\documentclass{iopjournal}

\usepackage[utf8]{inputenc}
\usepackage{textcomp}
\usepackage{amsmath}
\usepackage{booktabs}
\usepackage{tabularx}
\usepackage{array}
\usepackage[numbers,sort&compress]{natbib}
\usepackage{ragged2e}
\AtBeginDocument{\justifying}

\makeatletter
\renewcommand{\@makecaption}[2]{\vskip\abovecaptionskip
\sbox\@tempboxa{\fontsize{8}{10}\selectfont {\bfseries #1.} #2}%
\ifdim \wd\@tempboxa >\hsize
\justifying\noindent \fontsize{8}{10}\selectfont {\bfseries #1.} #2\par
\else
\global \@minipagefalse
\hb@xt@\hsize{\hfil\box\@tempboxa\hfil}%
\fi
\vskip\belowcaptionskip}
\makeatother

\begin{document}

\pagestyle{plain} 

\title{Probing the wave-vector gap in nanoscale magnonics}

\author{Ond\v{r}ej Wojewoda$^{1,2,*}$, Samer Kurdi$^3$,
Toeno van der Sar$^4$, Daniel Schick$^5$,
Sebastian Wintz$^6$, Alexei Maznev$^7$}

\affil{$^1$CEITEC BUT, Brno University of Technology, Purky\v{n}ova 123, 612 00 Brno, Czech Republic}
\affil{$^2$Department of Materials Science and Engineering, Massachusetts Institute of Technology, 02139 Cambridge, Massachusetts, United States of America}
\affil{$^3$Institute of Photonics and Quantum Sciences, SUPA, Heriot-Watt University, EH14 4AS Edinburgh, United Kingdom}
\affil{$^4$Department of Quantum Nanoscience, Kavli Institute of Nanoscience, Delft University of Technology, Lorentzweg 1, 2628 CJ Delft, The Netherlands}
\affil{$^5$Max Born Institute for Nonlinear Optics and Short Pulse Spectroscopy, Max-Born-Stra{\ss}e 2A, 12489 Berlin, Germany}
\affil{$^6$Helmholtz-Zentrum Berlin f\"ur Materialien und Energie, Albert-Einstein-Stra{\ss}e 15, 12489 Berlin, Germany}
\affil{$^7$Department of Chemistry, Massachusetts Institute of Technology, 02139 Cambridge, Massachusetts, United States of America}

\affil{$^*$Author to whom any correspondence should be addressed.}
\email{ondrej.wojewoda@vutbr.cz}

\keywords{}

\begin{abstract}
Short-wavelength spin waves are central to miniaturizing magnonic devices and to studying exchange-dominated magnetization dynamics. Yet the wave-vector range between roughly 100 and 300\,rad\,\textmu{}m$^{-1}$ has remained difficult to access experimentally, posing challenges for both excitation and detection. This Perspective reviews recent advances in emerging measurement techniques and discusses prospects for their future development. Optical and microwave techniques are restricted to the vicinity of the Brillouin-zone centre. Inelastic neutron, electron and resonant hard x-ray scattering cover the full Brillouin zone but lack the energy resolution needed for low-frequency magnons at intermediate wave vectors. This Perspective examines five emerging approaches that are beginning to close this gap: Mie-enhanced Brillouin light scattering, colour-centre magnetometry, magnon momentum microscopy, extreme-ultraviolet transient-grating spectroscopy, and time-resolved scanning transmission x-ray microscopy. For each, we discuss the physical principle, demonstrated capabilities, current limitations and prospects for development. Compared side by side, the techniques give complementary access to magnon frequencies, local stray fields, real-space dynamics and momentum-space distributions, and they differ in their sensitivity to coherent and incoherent excitations. We therefore discuss the main advantages and use cases of each technique and consider the practical requirements for addressing different scientific questions. We highlight opportunities for further development, including higher sensitivity, improved spatial and temporal resolution, and broader frequency and wavevector coverage. Together, these advances could enable, for example, systematic studies of nonlinear magnon interactions at nanoscale wavelengths and investigations of emerging magnetic materials such as altermagnets, while providing the tools needed to characterize nanoscale magnonic devices.
\end{abstract}

Spin waves are collective excitations of magnetic order, and magnons are their corresponding quanta. They are promising candidates for future computing and information-processing architectures because they transport information without charge motion, avoiding Ohmic losses during propagation, and have shorter wavelengths than electromagnetic waves at the same frequency. Their dispersion extends across the crystal Brillouin zone, with different interactions dominating at different length scales. For ferromagnetic materials, at shorter wavelengths, exchange interactions dominate, resulting in an approximately parabolic dispersion. At longer wavelengths, long-range dipolar (magnetostatic) interactions become important; in thin films, they strongly modify the dispersion, in certain geometries increase the magnitude of the group velocity, and enable efficient spatial magnon transport in this regime \cite{Chumak2015,Pirro2021,DamonEshbach1961,Kalinikos1986,Serga2010}. Both of these regimes have been addressed separately, with different sets of experimental probing techniques. Patterned microscale antennas coupled with electrical detection, and optical probes in real and reciprocal space, have enabled access to spin waves near the centre of the Brillouin zone \cite{Sebastian2015,Krcma2025}. Conversely, neutron, electron, and resonant inelastic hard x-ray scattering access momenta and energies across the whole Brillouin zone, but their energy resolution is insufficient for the low-frequency (GHz), small-to-intermediate wave-vector regime, and they offer limited or no spatial, temporal, or phase resolution.\cite{Krcma2025,Miedaner2024,Biesner2022,Vollmer2004,Ament2011}. However, in between these two approaches a gap emerged \cite{Krcma2025,Miedaner2024}.

The central subject of this Perspective is therefore filling the experimental blind spot between high-resolution techniques confined to the vicinity of the Brillouin-zone centre and scattering probes that access the full Brillouin zone. This gap spans approximately 30 to 300\,rad\,\textmu{}m$^{-1}$ at high frequencies (above 40\,GHz) and about 100 to 300\,rad\,\textmu{}m$^{-1}$ at low frequencies, depending on the method and excitation conditions \cite{Krcma2025}. This wave-vector range provides new opportunities for magnonics: many promising concepts have already been demonstrated on the microscale, but application feasibility, low-energy operation, and fast processing require further miniaturization. In wave-based computational devices, the device dimensions scale with the wavelength of the operating modes, so accessing shorter wavelengths is a prerequisite for miniaturization. Moreover, access to this previously inaccessible wave-vector range can provide new insight into fundamental physics, such as pinning of short-wavelength spin waves, nonlinear scattering, or chiral splitting in altermagnetic materials \cite{Krcma2025,Smejkal2023,Liu2024,Wittrock2026, Bittencourt2026}.

Optical techniques, primarily Brillouin light scattering (BLS), have been extensively used to probe spin waves. While originally BLS was used to measure spectra of thermal magnons, more recently microfocused BLS was developed to measure coherent spin waves generated by microwave antennas in magnonic devices \cite{Sebastian2015,Chumak2015,Pirro2021}. Optical microscopy techniques based on the magneto-optical Kerr and Faraday effects have also been used to visualize magnetic-field patterns of coherent spin waves excited either by microwave antennas or by ultrashort optical pulses \cite{Qin2021,Hashimoto2017}. In either case, the accessible magnon wave-vector range has an upper bound equal to twice the optical wave vector, \(k_{\max} = 2k_{\mathrm{light}} = 4\pi/\lambda_{\mathrm{light}}\), where \(\lambda_{\mathrm{light}}\) is the optical wavelength. (In other words, the magnon wavelength cannot be smaller than \(\lambda_{\mathrm{light}}/2\).) The emerging probes described here use different approaches to extend or circumvent this limit.

The most straightforward way to extend the wave-vector range is to reduce the optical wavelength by moving into the extreme ultraviolet (EUV) and soft-x-ray ranges, where scattering from spin waves is greatly enhanced at the M, N and L absorption edges of magnetic elements. Resonant inelastic x-ray scattering (RIXS) from spin waves is essentially an x-ray analogue of BLS capable of measuring magnon dispersion across the entire Brillouin zone, but the resolution of RIXS spectrometers, typically \(\sim\)20\,meV or \(\sim\)5\,THz \cite{Ament2011,Schlappa2024, Martinelli2022}, is many orders of magnitude poorer than that of the Fabry--Pérot spectrometers used in BLS; consequently, RIXS can only probe very high-frequency magnons with wave vectors greater than \(\sim\)300\,rad\,\textmu{}m$^{-1}$. However, the need for a high-resolution spectrometer can be circumvented if one deals with coherently excited rather than thermal magnons. For example, in soft-x-ray magnon momentum microscopy, a microwave antenna is used to generate spin waves at a fixed frequency, and their wave vector is measured via the soft-x-ray scattering angle \cite{Wittrock2026}. An alternative approach is to use soft-x-ray imaging to obtain information in real rather than reciprocal space: time-resolved x-ray microscopy provides real-space images of the phase and amplitude of nanoscale coherent spin waves \cite{Wintz2016,Trager2020}. While imaging yields richer information content than scattering measurements, the latter can potentially access higher wave vectors and frequencies, as the x-ray imaging resolution is still far from the diffraction limit and the time resolution is limited by the x-ray probe flash lengths. In another resonant quasi-elastic scattering method, referred to as EUV transient gratings \cite{Miedaner2024}, EUV radiation is used not only to probe but also to excite coherent spin waves: two crossed ultrashort EUV pulses produce standing spin waves at a set wave vector rather than at a set frequency. Resonant scattering of a time-delayed probe pulse is then measured as a function of the time delay, revealing the spin-wave oscillations. Thus, the measurement is done in reciprocal space to access the spin wave momentum and in time rather than the frequency domain, again circumventing the need for a spectrometer.

In an entirely different approach \cite{Krcma2025}, the wave-vector range of microfocused BLS is extended without resorting to short wavelengths of the probing light. Rather, a nanoscale Mie resonator fabricated on the sample adds the wave vectors provided by the light localization to the momentum-conservation relation. This approach extends the wave-vector range of microfocused BLS by an order of magnitude without any changes to the optical apparatus and can be applied to both thermal magnon spectroscopy and coherent spin waves generated in magnonic devices.

Finally, the restriction imposed by the optical wave vector can be entirely removed by using an atomic-scale probe interrogated optically. Colour-centre magnetometry detects the dipolar stray fields generated by spin waves \cite{Bertelli2020,ManasValero2026}. Nanometre-scale spatial resolution can be achieved with single-spin sensors -- typically a single nitrogen-vacancy (NV) centre integrated into a scanning probe and positioned close to the sample surface \cite{SimonKurdi2022}.

\begin{figure}
\centering
\includegraphics[width=\textwidth]{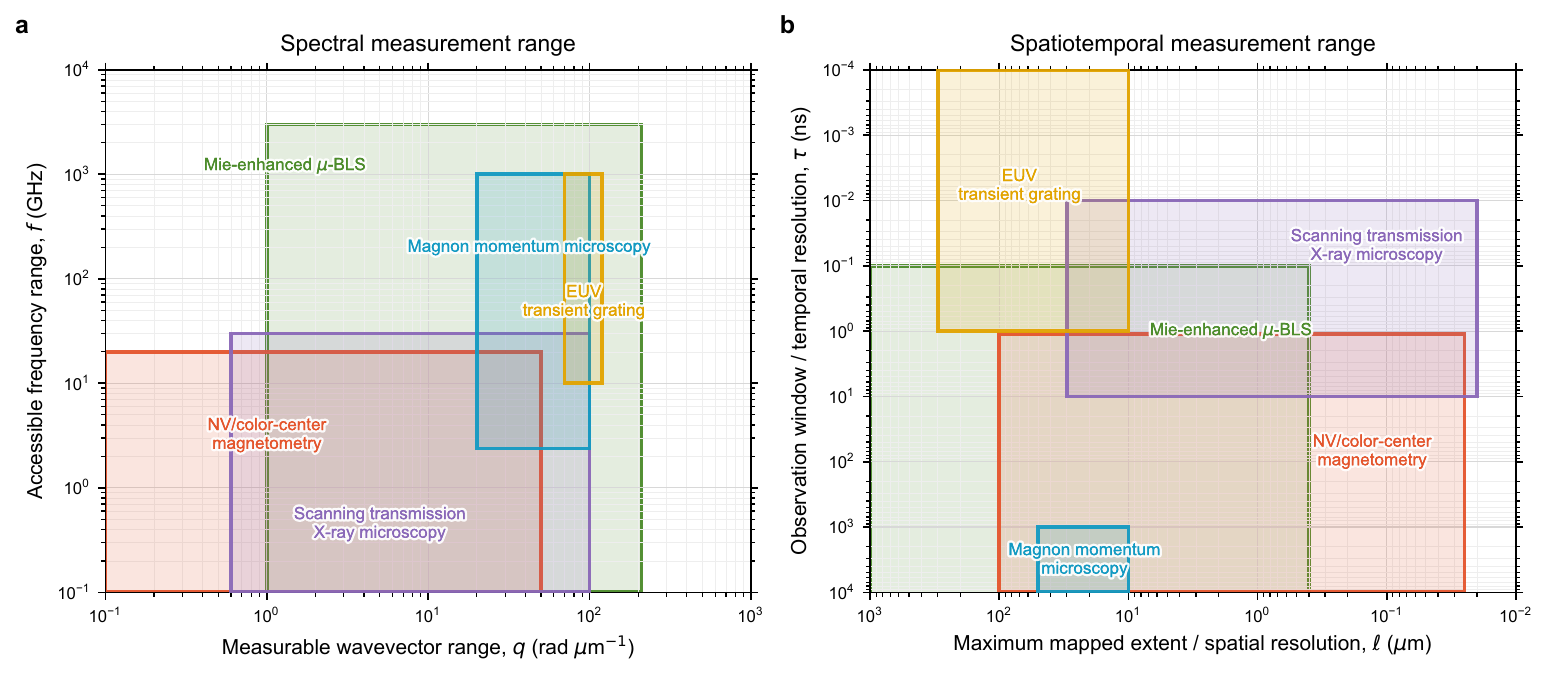}
\caption{\textbf{Capability landscape of the presented techniques.} Comparison of the demonstrated (a) accessible frequency and wavenumber ranges and (b) experimentally achievable temporal and spatial resolutions. For Mie-enhanced BLS, the maximum demonstrated wave vector is limited by the signal-to-noise ratio, while the accessible wave vectors are selected through the resonator periodicity and measurement-spot size. The frequency and temporal ranges are determined by the interferometer configuration, mirror spacing, and cavity finesse. The spatial resolution is set by the optical spot (\(\sim\)400\,nm), and the measurement positions by the resonator locations \cite{Krcma2025,Wojewoda2023phase}. For NV/colour-centre magnetometry, the shaded regions indicate representative practical operating ranges spanning different NV measurement modalities rather than fundamental limits. In the spectral domain (a), the accessible wave-vector range depends on the sensor--sample-distance filter function. The accessible frequency range combines different resonant and non-resonant detection modalities and should not be interpreted as the continuous bandwidth of a single protocol. Broadband NV (A)FMR detection has been demonstrated up to approximately 20--24\,GHz \cite{Melendez2025}, while resonant detection can be shifted by an applied magnetic field through the Zeeman interaction. In the spatiotemporal domain (b), scanning NV magnetometry has demonstrated a magnetic stand-off distance as small as approximately 25\,nm \cite{Xu2025}, while widefield imaging is optically diffraction limited (500\,nm). Temporal resolutions down to 1.1\,ns have been demonstrated for transient-field sensing \cite{Herb2025} but not for spin-wave imaging. For EUV transient-grating spectroscopy, the accessible wave vector is determined by the grating period, while the frequency range is limited by the pump--probe delay range and pulse duration. The temporal resolution is approximately 100\,fs; the technique does not provide real-space imaging \cite{Miedaner2024}. For magnon momentum microscopy, the accessible wave-vector range is generally limited by the diffraction limit, potentially reaching \(>\)1000\,rad\,\textmu{}m$^{-1}$ for typical soft-x-ray wavelengths. Further, there is no intrinsic frequency limitation within the relevant GHz--THz range. The present quasi-elastic implementation has no temporal resolution, although femtosecond-to-picosecond resolution may be possible in a pump--probe configuration; the spatial resolution and spatial extent are determined by the micrometre-scale beam footprint, which can easily be increased, albeit at the expense of resolution in reciprocal space \cite{Wittrock2026}. For scanning transmission x-ray microscopy, the wave-vector and spatial resolutions are determined by the x-ray spot size (\(\sim\)25\,nm), the accessible frequency range by the microwave electronics, and the temporal resolution by the x-ray pulse duration (\(\sim\)10\,ps in low-alpha operation) \cite{Mayr2024}.}
\label{fig:landscape}
\end{figure}

These approaches are complementary, as each offers a different balance between accessible wave-vector and frequency ranges, spatial and temporal resolution, and sensitivity to coherent versus incoherent magnons, as summarized in figure~\ref{fig:landscape}. In the following sections, we discuss each technique separately (Mie-enhanced BLS, colour-centre magnetometry, magnon momentum microscopy, EUV transient gratings, and scanning transmission x-ray microscopy (STXM)), outlining its underlying physical principle, its current limitations, and the outlook for further development and applications. We conclude with a side-by-side comparison of the techniques (table~\ref{tab:comparison}).

\section{Mie-enhanced Brillouin light scattering}\label{sec:mie}
\begin{figure}
\centering
\includegraphics[width=\textwidth]{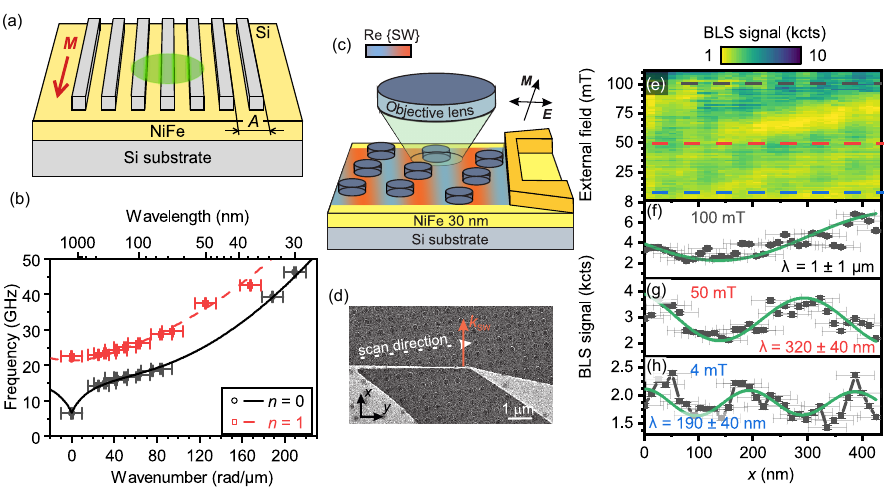}
\caption{\textbf{Examples of Mie-enhanced BLS measurements.} (a), (b) Schematic (a) and extracted dispersion relation (b) obtained by Mie-enhanced $k$-resolved BLS at 50\,mT in a permalloy thin film. Reproduced from \cite{Krcma2025}. (c), (d) Schematic (c) and scanning electron micrograph (d) of a phase-resolved BLS measurement of coherently excited spin waves in a 30\,nm NiFe (permalloy) thin film. The white line in (d) indicates the scan direction. (e)--(h) Interference signal of coherently excited spin waves measured along the line marked in (d). Panel (e) shows the full bias-field sweep, while (f)--(h) show line profiles measured at 100\,mT, 50\,mT, and 4\,mT, respectively. Reproduced from \cite{Wojewoda2023phase} (CC BY 4.0).}
\label{fig:mie}
\end{figure}
Brillouin light scattering (BLS), first described by Léon Brillouin, is a well-established technique based on the inelastic scattering of visible light \cite{Brillouin1922,Merklein2022}. A coherent laser beam illuminates the sample, and the backscattered light is collected and spectrally analysed, typically using a tandem Fabry--Pérot interferometer or a virtually imaged phased array (VIPA) \cite{Lindsay1981,ChagnonLessard2026}. During scattering, a magnon is created (Stokes process) or annihilated (anti-Stokes process), shifting the photon frequency by the magnon frequency. Momentum conservation requires the magnon wave vector to equal the momentum transferred between the incident and scattered photons. Consequently, BLS is limited to \(k_{\max} = 2k_{\mathrm{light}}\); for 532\,nm light, this corresponds to \(k_{\max} \approx 23.6\)\,rad\,\textmu{}m$^{-1}$, or a minimum spin-wave wavelength of approximately 266\,nm \cite{Wojewoda2023,Wojewoda2024Modeling}.

Conventional, or $k$-resolved, BLS enabled early measurements of spin-wave spectra in thin magnetic films and multilayers \cite{Sandercock1979,Grimsditch1979,Sebastian2015,Madami2012}. Advances in micro- and nanofabrication allowed for smaller magnetic structures, which drove the development of microfocused BLS (\textmu{}BLS) \cite{Sebastian2015,Wojewoda2024Modeling}. This approach trades wave-vector selectivity for diffraction-limited real-space resolution, allowing the local spin-wave intensity to be mapped in patterned structures. It can also be extended to time- and phase-resolved measurements \cite{Sebastian2015,Wojewoda2023phase}. Nevertheless, focusing alone does not overcome the optical momentum limit and therefore cannot enable measurements of spin waves with wavelengths shorter than those accessible to conventional BLS.

This limit can be overcome using nanophotonic resonators with dimensions smaller than the diffraction limit of the probing light, i.e., below \(\lambda_{\mathrm{light}}/2\). Their localized near fields contain wave-vector components exceeding \(k_{\mathrm{light}}\), thereby supplying the additional momentum required to probe spin waves with \(k > k_{\max}\) \cite{Wojewoda2023}. Early implementations employed metallic nanostructures supporting plasmonic resonances; although these structures extended the accessible wave-vector range, optical losses and heating strongly reduced the usable BLS signal \cite{Freeman2020,Lozovski2024,Jersch2010}. High-index dielectric resonators provide a low-loss alternative: their Mie resonances simultaneously broaden the wave-vector sensitivity and enhance the detected signal \cite{Wojewoda2023,Wojewoda2023phase,Krcma2025}. 

Since the whole ecosystem of \textmu{}BLS is already established, Mie resonators can take full advantage of all the available techniques and even introduce new possibilities, such as wave-vector-resolved measurements. This was demonstrated on a thin NiFe film on which periodic dielectric nanoresonators were patterned (figure~\ref{fig:mie}(a)) \cite{Krcma2025}. By measuring on different periodicities, one can retrieve the whole dispersion relation (figure~\ref{fig:mie}(b)).

However, as already mentioned, Mie resonances can also be used with coherently excited spin waves. In the experiment presented in figures~\ref{fig:mie}(c) and (d), the individual resonators were fabricated as a tilted array near the excitation antenna \cite{Wojewoda2023phase}. By using a standard approach with a constant-phase signal, the phase of the spin waves can be retrieved. This measurement can be used to measure not only the dispersion relation but also effects such as phase shifting or the coupling of spin waves that are otherwise inaccessible to free-space light (wavelengths below approximately 250\,nm).

\subsection{Limitations}\label{mie:limitations}

Although Mie-enhanced BLS removes the upper wave-vector limit imposed by far-field momentum conservation, this gain does not translate into improved two-dimensional spatial resolution. The technique can detect spin waves with subdiffractional wavelengths, but real-space information remains limited by the optical probe geometry and by the spatial distribution of the resonators. Therefore, in its current implementation, Mie-enhanced BLS is not suitable for subdiffraction-limited imaging.

A second limitation is that Mie-enhanced BLS requires resonators to be nanofabricated on the sample before the measurement. These structures define in advance where the near-field enhancement occurs, and therefore where the spin-wave signal can be efficiently detected. This fixed measurement geometry restricts the flexibility of the technique and limits its applicability to samples or device layouts that can tolerate prior nanofabrication.

\subsection{Perspective}\label{mie:perspective}

Although Mie-enhanced BLS has now been experimentally demonstrated in several measurement configurations, substantial development is still needed to fully exploit its potential.

Further gains may be achieved by optimizing the resonator geometry and material, analogous to the development of near-field Raman enhancement, where enhancement factors increased from below 10\textsuperscript{2} to as high as 10\textsuperscript{12} \cite{Jeanmaire1977,LeRu2007}. Higher-order Mie modes, anapole resonances, three-dimensional resonator designs, and lower-loss materials could improve both wave-vector sensitivity and signal intensity \cite{Wojewoda2023}. Such advances could extend BLS towards deeply sub-10\,nm spin-wave wavelengths and potentially enable studies of spin-wave propagation in atomically thin magnets.

Benchmarking Mie-enhanced BLS with coherently excited short-wavelength spin waves could reveal its ultimate wave-vector range, which is difficult to assess using weak thermal-magnon signals alone. Novel coherent sources, including antenna-coupled spin textures, grating couplers, and parametrically pumped spin waves \cite{Hollander2018,Yu2013,Heinz2022}, would provide well-defined high-wave-vector excitations for this purpose. Moreover, because BLS provides intrinsic frequency resolution, Mie-enhanced BLS could become a powerful tool for studying nonlinear processes in which spin waves scatter into short-wavelength states.

Finally, integrating a Mie resonator onto a scanning probe could remove the need to fabricate nanostructures directly on each sample. Inspired by tip-enhanced Raman microscopy and aperture-based near-field BLS, such an approach could combine subdiffractional imaging with enhanced signal levels \cite{Jersch2010,Verma2017}. A reusable and optimized probe would accelerate spin-wave-device prototyping and extend Mie-enhanced BLS to samples that are incompatible with direct nanofabrication, including biological systems.

\section{NV/colour-centre magnetometry}\label{sec:nv}

The negatively charged nitrogen-vacancy (NV) centre in diamond is an optically addressable spin defect whose electron-spin-resonance (ESR) transitions provide a local probe of dynamic magnetic fields \cite{Casola2018}. Coherent spin waves are detected by positioning a shallow NV centre close to the magnetic material and tuning an NV transition into resonance with the spin-wave frequency. The spin-wave stray field is combined coherently with a phase-stable microwave reference field, producing a spatial interference pattern as the phase of the propagating spin wave evolves relative to the reference \cite{Bertelli2020,Zhou2021}. This interference can be read out either through the modulation of the NV Rabi frequency or through changes in the ESR contrast. In Rabi measurements, the local transverse microwave-field amplitude is obtained from the measured Rabi frequency, enabling quantitative measurements of the spin-wave amplitude. In NV-ESR contrast measurements, the interference modifies the depth of the ESR contrast dip, providing a fast, quantitative spatial map of the coherent spin-wave field.

Early experiments demonstrated ESR-resonant detection of spin waves in a permalloy microdisc \cite{VanDerSar2015} and detection over millimetre-scale distances in an yttrium iron garnet (YIG) disc \cite{Kikuchi2017}. Subsequent confocal ensemble measurements enabled spatially resolved imaging of coherent spin-wave propagation, decay, interference and scattering \cite{Bertelli2020,Zhou2021}.

\begin{figure}[htbp]
\centering
\includegraphics[width=\textwidth]{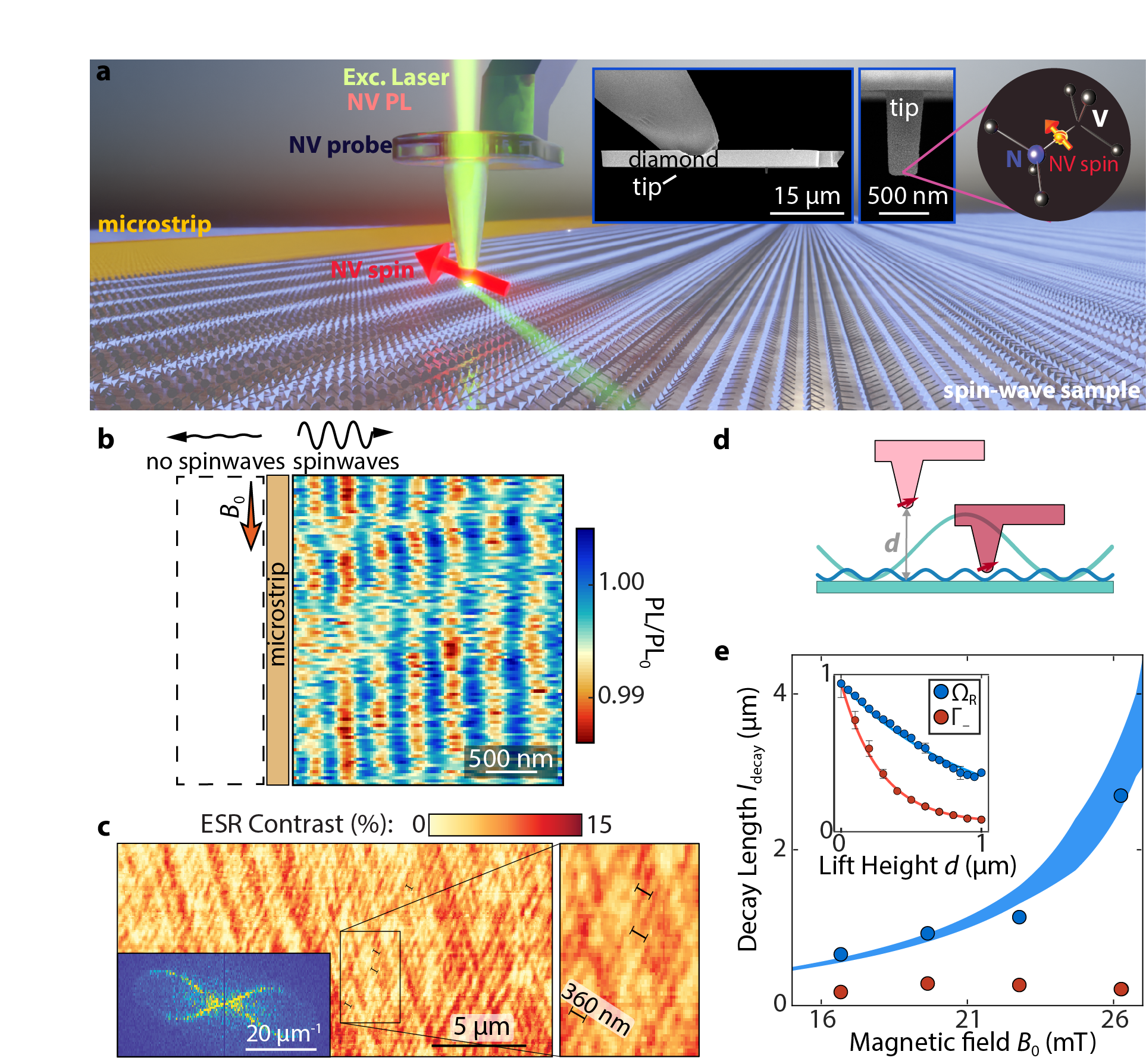}
\caption{\textbf{Nanoscale probing of spin waves using nitrogen-vacancy spin(s) in diamond.} (a) Schematic of a typical scanning nitrogen-vacancy (NV) experiment on spin waves. A single NV centre located approximately 20\,nm below the apex of a diamond tip detects the magnetic stray fields generated by spin waves excited in a thin-film magnet, e.g. yttrium iron garnet (YIG), using a microwave microstrip. The NV spin is optically initialized and read out with a green laser through spin-dependent photoluminescence (PL), while scanning-probe feedback maintains the tip--sample separation. Reproduced with permission from \cite{SimonKurdi2022}. Copyright (2022) American Chemical Society. Insets show a home-made diamond tip (reproduced from \cite{Simon2021}; copyright (2021) American Chemical Society) and the orientation of the NV spin (reproduced from \cite{Carmiggelt2023} with permission from Springer Nature). (b) Spatial map of the normalized NV electron-spin-resonance PL signal (PL/PL\textsubscript{0}), revealing 500\,nm unidirectionally excited coherent spin waves. (c) ESR-contrast (\(1-\)PL/PL\textsubscript{0}) spatial map showing the spatial interference pattern produced by spin-wave scattering in a YIG film. The inset presents the corresponding Fourier-transform map, revealing the wave-vector distribution via an isofrequency contour of the spin-wave dispersion. (d) Schematic of the exponential decay of the spin-wave stray field with increasing NV--sample separation \(d\). Tuning \(d\) optimizes the detection sensitivity for different wavelengths. (e) Demonstration of (d) via measurements of the spin-wave decay length \(l_{\mathrm{decay}}\) as a function of the magnetic field \(B_{0}\) for driven coherent spin waves (blue) and a magnon gas (red). The solid blue line and shaded region show the calculated decay length and its uncertainty for coherent spin waves resonant with the NV ESR transition. The inset shows representative lift-height measurements of the NV Rabi frequency (\(\Omega_{\mathrm{R}}\)) for coherently driven modes and of the NV relaxation rate (\(\Gamma_{-}\)) for the magnon gas at \(B_{0} = 20\,\mathrm{mT}\), with solid lines denoting exponential fits used to extract the decay lengths. Panels (b, e) and (c, d) are reproduced from Refs.~\cite{Simon2021} and \cite{SimonKurdi2022}, respectively. Copyright (2021) and (2022) American Chemical Society, respectively.}
\label{fig:nv}
\end{figure}

The transition towards nanoscale spin-wave imaging was enabled by advances in scanning-probe NV magnetometry, which evolved from nanodiamonds grafted onto scanning tips \cite{Grinolds2013,Cuche2009} to microfabricated all-diamond nanopillars with improved spin coherence, photon collection, and control of the sensor--sample spacing \cite{Maletinsky2012}. Initially, ensemble scanning tips were utilized to probe directional surface spin waves with wavelengths of \(\sim\)500\,nm \cite{Simon2021} (figures~\ref{fig:nv}(a) and (b)); later, with single-spin probes, scanning NV magnetometry enabled imaging of short-wavelength spin waves in YIG \cite{SimonKurdi2022} (figures~\ref{fig:nv}(a) and (c)). The probe height also provides an additional degree of wave-vector selectivity because the stray field of a spin-wave mode with wave vector \(k\) decays approximately as \(\mathrm{e}^{-kd}\), where \(d\) is the NV--sample distance (figure~\ref{fig:nv}(d)). Lift-height-dependent measurements can therefore be used to distinguish contributions from different wave-vector components \cite{Simon2021} (figure~\ref{fig:nv}(e)), while maintaining \(kd\) approximately constant preserves strong imaging contrast over a broad range of spin-wave wavelengths \cite{SimonKurdi2022}.

Despite these advances, direct detection of coherent spin-wave stray fields remains restricted to frequencies near the NV ESR transition. Several complementary approaches have been developed to expand the accessible frequency range, as outlined below:

\begin{enumerate}
\renewcommand{\labelenumi}{\arabic{enumi})}
\item
  \textbf{Off-resonant detection:} When the spin-wave frequency is detuned from the NV ESR transition, off-resonant detection schemes have been used and demonstrated in YIG. One approach measures the excitation-induced reduction in the longitudinal magnetization upon spin-wave excitation \cite{VanDerSar2015}. Another approach uses spin-wave down-conversion to NV-accessible MHz frequencies \cite{Carmiggelt2023}. These two approaches could enable spin-wave detection up to tens of GHz away from the NV ESR frequency, with improved sensitivity possible using alternating-current (AC) detection techniques \cite{Maze2008}. Coherent detection over a finite bandwidth around the NV resonance can additionally be achieved through four-spin-wave mixing, currently demonstrated up to 1\,GHz beyond the NV ESR frequency \cite{Carmiggelt2023}.
\item
  \textbf{Isofrequency imaging:} By selecting the sensor anisotropy axis orthogonal to the magnetization of the magnetic film -- for example, by using an out-of-plane-oriented (111) NV centre and aligning the film magnetization in the plane -- the sensor ESR frequency becomes, to first order, insensitive to variations in the applied field, while the spin-wave dispersion remains field-tunable \cite{ManasValero2026}.
\item
  \textbf{Broadband (anti)ferromagnetic resonance detection:} Driving (anti)ferromagnetic resonance ((A)FMR) modes can generate a broadband population of scattered magnons, resulting in increased magnon densities at the NV ESR frequencies that cause NV-spin relaxation. This relaxation process is observed as a broadband NV-photoluminescence reduction \cite{Wolfe2016} and was subsequently used to measure the magnon chemical potential \cite{Du2017} and the directional excitation of dense nonequilibrium magnon gases \cite{Simon2021}. More recently, the same principle enabled optical detection of antiferromagnetic-resonance modes at frequencies up to 24\,GHz in bulk van der Waals (vdW) magnets, far above the intrinsic NV resonance frequency \cite{Melendez2025}.
\end{enumerate}

More recently, boron-vacancy centres (V\textsubscript{B}\textsuperscript{$-$}) in hexagonal boron nitride (hBN) have emerged as a complementary colour-centre sensing platform: their spin-resonance spectrum and two-dimensional host geometry extend the accessible spin-wave frequency range and permit defect ensembles to be placed directly onto magnetic films or integrated into vdW magnet heterostructures, enabling coherent spin-wave imaging \cite{ManasValero2026}. Recent results show promise towards coherent magnon imaging at millikelvin temperatures \cite{Zhou2026}.

\subsection{Limitations}\label{nv:limitations}

A key challenge for colour-centre magnetometry in nanoscale spin-wave imaging is the limited frequency range over which coherent dynamics can be detected with high sensitivity. Resonant measurements are generally restricted to frequencies near the sensor ESR transitions, whereas many spin-wave modes of interest lie several gigahertz away. Although substantial progress has been made in off-resonant detection, coherent imaging of nanoscale spin waves far detuned from the ESR frequency remains elusive. This limitation is particularly important for vdW magnets, where strong magnetic anisotropy can shift the spin-wave spectrum above 10\,GHz and beyond the most accessible NV-detection window \cite{GrubisicCabo2025}. Extending the sensing bandwidth while preserving nanoscale spatial resolution is therefore essential for imaging propagating spin waves in these materials.

A second limitation is the NV--sample separation, which sets both the spatial resolution and the sensitivity to short-wavelength spin waves. Although state-of-the-art scanning probes can host NV centres within 10\,nm of the diamond apex, the effective magnetic stand-off commonly remains between 35 and 120\,nm owing to surface protrusions, adsorbates and imperfect mechanical contact \cite{Xu2025}.

For hBN magnetometry, state-of-the-art magnetic-field sensing has primarily relied on the negatively charged boron-vacancy defect V\textsubscript{B}\textsuperscript{$-$}, with spin \(S=1\). Owing to its low quantum efficiency \cite{Vogl2026}, measurements have so far been restricted largely to defect ensembles, limiting the spatial resolution to the optical diffraction limit.

\subsection{Perspective}\label{nv:perspective}

Colour-centre magnetometry is developing rapidly as a platform for nanoscale spin-wave imaging; however, several key advances are still required to extend its capabilities. We outline several directions below.

First, minimizing the sensor--sample distance is essential for accessing exchange-dominated, short-wavelength spin waves. Improved tip fabrication, surface preparation and frequency-modulated force feedback should enable more reliable stand-off distances below 10\,nm \cite{Xu2025}, corresponding to sensitivity to wave vectors of the order of 100\,rad\,\textmu{}m$^{-1}$. Single-spin hBN sensors could reduce this distance further, provided that recently identified carbon-related defects can be developed into reproducible probes \cite{Gilardoni2025}.

Second, direct imaging of ESR-resonant modes at higher frequencies will require broadband microwave delivery together with large magnetic fields. For example, accessing frequencies of 20--40\,GHz requires fields of approximately 0.6--1.3\,T and 0.8--1.5\,T for the \(m_{s} = 0 \rightarrow + 1\) and \(m_{s} = 0 \rightarrow - 1\) transitions, respectively, which are readily achievable using laboratory electromagnets and superconducting magnets. Complementary time-domain protocols could remove the need for a spatially uniform RF reference field. The recent demonstration of transient NV magnetometry with a temporal resolution approaching 1\,ns \cite{Herb2025} could enable direct imaging of propagating spin waves in the sub-gigahertz regime.

Third, magnetic sensing also provides access to spin waves beneath optically opaque structures. This capability could be used to image wavelength conversion and confinement produced by metallic Fabry--Pérot resonators \cite{Qin2021}, antidot structures \cite{Gross2020} and superconducting elements \cite{Borst2023}. Colour-centre magnetometry could therefore reveal how functional overlayers modify the local dispersion, phase and propagation direction of nanoscale spin waves.

Fourth, millikelvin scanning NV magnetometry \cite{Scheidegger2022} could extend nanoscale spin-wave imaging to low temperatures. Removing free-space optics, for example via fibre-coupled probes, may reduce the optical power required in dilution refrigerators \cite{Li2024,Li2023}, providing potential for low-temperature magnon transport.

Finally, combining colour-centre magnetometry with differential-reflectivity measurements could correlate the magnetic stray field of a mode with its coupling to excitons, as investigated most prominently in the vdW magnet CrSBr, providing complementary information on magnon propagation and hybrid magnon--exciton interactions \cite{Bae2022}.

\section{Magnon momentum microscopy}\label{sec:mmm}

Spin waves can be described as a spatially extended, periodic modulation of the magnetization in a material. When this modulation can be \emph{seen} by light, it can act as a grating, leading to well-defined diffraction spots. The intensity and position of the diffraction signal in reciprocal space directly encode information on the spin wave's wavelength, direction, and amplitude. In this scenario, using soft x-rays as a probe has two main advantages: (i) the magnetic contrast at transition-metal L edges and rare-earth M edges is extremely large \cite{Kortright2013}; (ii) the few-nanometre wavelength of soft x-rays provides access to high-wave-vector spin waves (see figure~\ref{fig:landscape}(a)).

Since magnon momentum microscopy (MMM) in its basic implementation only resolves the finite momentum of spin waves, the scattering process can be assumed to be quasi-elastic, as the energy transferred between the magnons and photons (10--100\,\textmu{}eV) is negligible compared with the \(\sim\)1\,keV photon energy in the soft x-ray regime. Efforts to actually resolve this energy transfer in resonant inelastic x-ray scattering (RIXS) -- the x-ray equivalent of Brillouin light scattering -- require extremely sophisticated and photon-hungry spectrometers at synchrotrons or free-electron lasers (FELs). Alternative approaches working in time and real space rather than in reciprocal space, such as time-resolved microscopy (see section~\ref{sec:stxm}), offer the great advantages of high spatial resolution and real-space information, but are fundamentally constrained by their temporal and spatial resolution in accessing high-frequency and large-wave-vector spin waves.

Especially for the study of coherently excited spin waves, e.g. by a coplanar-waveguide antenna, the sensitivity of MMM increases with the probed volume in which the spin waves propagate. The detectable signal is then dominated by the simple absorption contrast of the magnetic grating due to x-ray magnetic circular dichroism (XMCD) \cite{Schutz1987,vanderLaan1986}. In addition to this very high sensitivity, MMM further enables a large region of momentum space to be recorded simultaneously without scanning.

\begin{figure}[htbp]
\centering
\includegraphics[width=\textwidth]{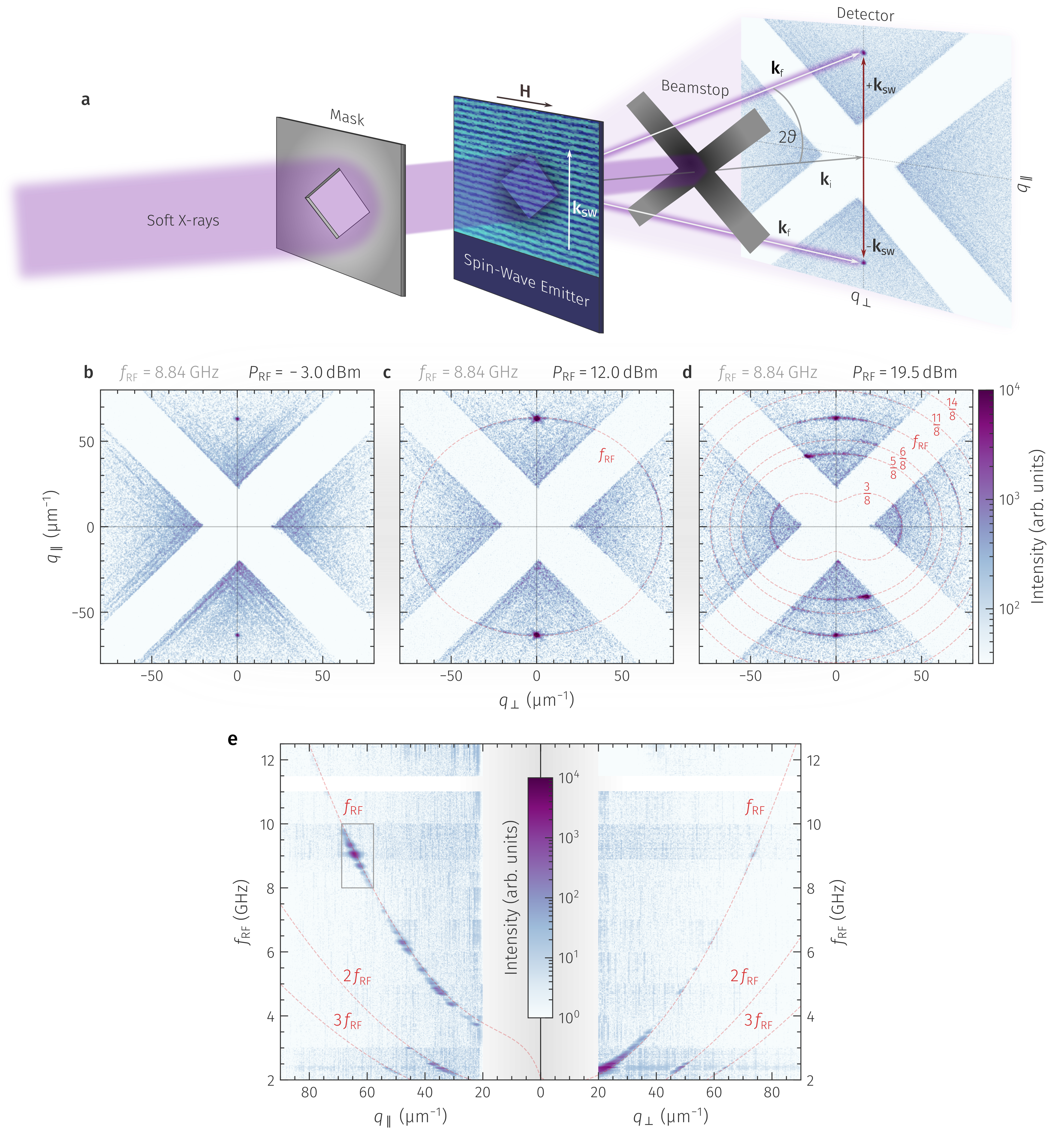}
\caption{\textbf{Magnon momentum microscopy for probing high-wave-vector magnons.} (a) Experimental setup showing the YIG sample, including the spin-wave emitter, as well as the mask and beamstop for the incoming soft x-rays. Two diffraction spots of +1 and \(-1\) order are directly observable on the area detector. (b)--(d) Hierarchy of nonlinear magnon scattering processes for increasing RF excitation power. (e) Dispersion relation of magnon modes in the Damon--Eshbach (DE; left) and backward-volume-wave (BVW; right) configurations. Figure reproduced and adapted from Ref.~\cite{Wittrock2026} (CC BY 4.0).}
\label{fig:mmm}
\end{figure}

The first experimental implementation of MMM is shown in figure~\ref{fig:mmm}(a) and is very similar to a standard small-angle x-ray scattering setup \cite{Wittrock2026}. Here, magnons were generated in a 100\,nm thick YIG film on a gadolinium gallium garnet (GGG) substrate by a coplanar waveguide and a nanoscale grating coupler in the Damon--Eshbach geometry. MMM detection was carried out at the Fe L\textsubscript{3} resonance near 708\,eV. A defined x-ray interaction region and a beamstop suppress parasitic scattering from the patterned antenna structures and saturation of the detector by the direct beam, resulting in a quasi-background-free magnetic diffraction signal, as shown in figure~\ref{fig:mmm}(b).

The most distinctive capability of MMM becomes apparent when the spin-wave population departs from a single well-defined mode. With increasing RF excitation power of the microwave antenna, the two distinct diffraction peaks in figure~\ref{fig:mmm}(b) broaden into an extended elliptical ring in momentum space as nonlinearly excited modes are populated (figure~\ref{fig:mmm}(c)). This directional population of magnon modes in the backward-volume geometry can be attributed to a parametric four-magnon process in which two driven magnons scatter into two secondary magnons of the same frequency but different propagation directions. This process is very similar to the second-order Suhl instability but at the same time distinct from it, as it is based on two initially propagating magnons rather than \(k=0\) modes. At even higher excitation powers (see figure~\ref{fig:mmm}(d)), the population of fractional harmonics with a complex azimuthal intensity profile clearly visualizes the nonlinear mode coupling in the sub-100\,nm wavelength regime accessible by MMM.

The same principle allows the spin-wave dispersion to be reconstructed by repeating the measurement over the excitation frequency. In the demonstrated system, the resulting momentum maps resolve both Damon--Eshbach and backward-volume propagation directions and reveal the approximately parabolic dispersion expected once exchange interactions dominate (figure~\ref{fig:mmm}(e)). The measurement can therefore bridge the experimentally difficult region between long-wavelength magnetostatic modes and the much larger momenta normally accessed by inelastic scattering techniques such as RIXS. Moreover, because the full two-dimensional momentum distribution is recorded, deviations from simple single-mode excitation -- such as mode competition, nonlinear scattering, or population of otherwise weakly excited branches -- become directly observable.

\subsection{Limitations}\label{mmm:limitations}

The advantages of MMM arise from its quasi-elastic nature, but this is also its most fundamental limitation. The technique does not intrinsically resolve the energy of the scattered photon and therefore does not provide a direct measurement of the magnon energy or frequency. In the present implementation, the magnon frequency is defined by the external microwave excitation, and the measured scattering pattern gives the corresponding momentum distribution. MMM is consequently complementary to energy-resolved techniques such as Brillouin light scattering and RIXS.

The momentum range shown in figure~\ref{fig:mmm}(e) should likewise not be interpreted as being limited by the demonstrated 67\,nm wavelength. At 708\,eV, the soft-x-ray wavelength is approximately 1.75\,nm, corresponding to a fundamental diffraction limit above 7000\,rad\,\textmu{}m$^{-1}$. The experimentally accessible range is instead constrained by the detector geometry and, at small momentum transfer, by the beamstop. In the present setup, scattering below approximately 20\,rad\,\textmu{}m$^{-1}$ is obscured. Both limits are principally geometrical and can be modified through the detector distance, beamstop design and scattering geometry.

Another limitation, shared with EUV transient-grating spectroscopy and STXM (see Sections~\ref{sec:euvtg} and~\ref{sec:stxm}), is that MMM requires EUV or soft x-ray sources typically available only at large-scale facilities, such as synchrotrons or XFELs. In the transmission geometry presented here, samples must also transmit a substantial fraction of the incident soft x-rays, requiring substrate back-thinning for samples that can only be grown on crystalline substrates. The backthinning also limits the spatial extent of the x-ray region. The demonstrated experiment uses a micrometre-scale illuminated area, which provides sufficient scattering volume for highly sensitive detection but averages over the illuminated region. This is advantageous for probing extended spin-wave populations but becomes less suitable when the relevant dynamics are confined to nanoscale device elements or when propagation lengths are short. Smaller apertures or tightly focused x-ray beams could reduce the interaction volume towards the submicrometre regime, although the resulting reduction in photon flux would need to be balanced against the desired spatial selectivity.

\subsection{Perspective}\label{mmm:perspective}

The most compelling future applications of MMM arise where the \emph{distribution} of magnons in momentum space is itself the quantity of interest. Nonlinear magnonics is an obvious example. Parametric processes, four-magnon scattering, and mode conversion redistribute populations across the dispersion plane, making a single-mode measurement intrinsically incomplete. A two-dimensional momentum map can instead reveal which regions of reciprocal space become populated. The observation of omnidirectional secondary modes in YIG provides an initial demonstration of this capability, but more complex magnetic systems are likely to produce substantially richer momentum-space signatures.

The combination of momentum resolution with element specificity further opens MMM to multilayers and heterogeneous magnetic systems. Resonant soft-x-ray scattering can selectively probe different chemical constituents, potentially allowing coupled spin-wave modes in multilayers to be studied separately. The bulk sensitivity of the technique also distinguishes it from surface-sensitive probes and makes buried magnetic functionality accessible without requiring the active magnetic layer to be exposed at the sample surface.

Reducing the x-ray interaction volume should extend MMM towards increasingly confined magnonic structures. The present micrometre-scale geometry is well suited to extended films, but focused soft-x-ray beams could enable measurements of individual device elements and systems with short spin-wave propagation lengths. Such an approach would bring reciprocal-space detection closer to the length scales of emerging magnonic devices, while retaining the ability to observe the complete momentum distribution rather than a single selected mode.

A particularly promising direction is the extension of MMM towards antiferromagnetic and other high-frequency magnetic systems. Because the detection principle is not intrinsically tied to microwave frequencies, the method could in principle be extended from GHz towards the THz regime. The challenge is to establish sufficient resonant magnetic contrast for systems with little or no static net magnetization. XMCD based on a dynamical net magnetization \cite{Blank2023}, or x-ray magnetic linear dichroism (XMLD), may provide suitable contrast, depending on the magnetic symmetry and experimental geometry. Accessing the momentum distribution of THz exchange magnons would provide a direct complement to energy-resolved techniques that are already capable of reaching large portions of the Brillouin zone but are comparatively inefficient for low-energy excitations.

Time-resolved implementations could add a further dimension to this reciprocal-space picture. Combining MMM with pulsed soft-x-ray sources would allow the evolution of a magnon population to be followed after an ultrafast excitation, potentially resolving the generation, propagation and nonlinear redistribution of spin waves in both momentum and time. In this context, laboratory-based high-harmonic-generation and plasma sources are particularly attractive because their short pulses and flexible repetition rates could make time-resolved momentum microscopy accessible beyond large-scale x-ray facilities. Such experiments could connect ultrafast changes in magnetic order directly to the subsequent population of finite-wave-vector modes.

Ultimately, MMM should be viewed not as a replacement for real-space or energy-resolved magnon probes, but as a complementary reciprocal-space observable. Its particular strength is the combination of high sensitivity, direct two-dimensional momentum access and element-specific magnetic contrast. This makes it especially well suited to phenomena in which magnons occupy a broad region of reciprocal space. As short-wavelength magnonics moves towards increasingly complex materials and device architectures, direct observation of this momentum-space population may become as important as measuring the dynamics of individual modes.

\section{EUV transient grating}\label{sec:euvtg}

\begin{figure}[htbp]
\centering
\includegraphics[width=\textwidth]{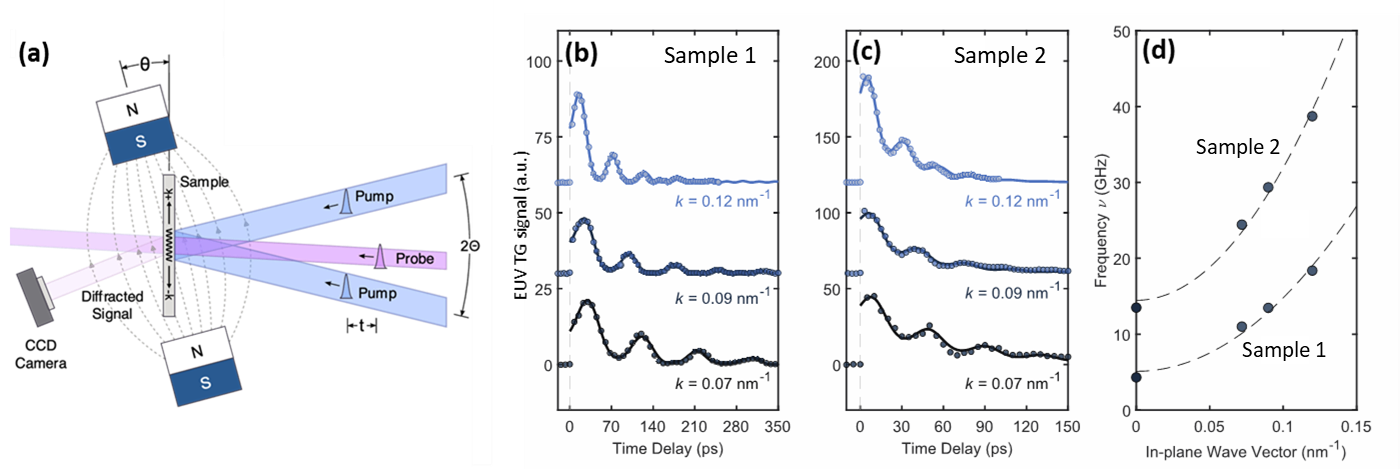}
\caption{\textbf{EUV TG magnon spectroscopy on Fe/Gd ferrimagnetic multilayer samples.} (a) Experimental setup. Two crossed EUV pump pulses launch counter-propagating magnons in the sample. A time-delayed EUV probe pulse is diffracted by the spatially periodic modulation of the magnetization, yielding a signal on the CCD camera. A tilted external magnetic field facilitates launching magnetic precession following ultrafast demagnetization at the grating maxima. (b) and (c) Diffracted signal as a function of the pump--probe delay for the two samples with slightly different Fe/Gd ratios. (d) Measured magnon dispersion curves for the two samples. Dashed curves represent parabolic fits. The figure is adapted from Ref.~\cite{Miedaner2024} (CC BY 4.0).}
\label{fig:euvtg}
\end{figure}

In contrast to the other methods described in this article, EUV transient grating (TG) spectroscopy is an `all-optical' technique, in which short-wavelength radiation is used not only to detect coherent magnons, as in MMM and STXM, but also to generate them. In the TG technique (see figure~\ref{fig:euvtg}(a)), the interference of two laser pulses crossed at the sample results in an interference pattern with a period of \(L = \lambda/(2\sin\Theta)\), where \(\lambda\) is the optical wavelength and \(2\Theta\) is the crossing angle. The interaction of the radiation with the sample leads to a spatially periodic material excitation with a period equal to $L$. The dynamics of this excitation are monitored via diffraction of a time-delayed probe pulse: the periodic excitation acts as a transient diffraction grating. TG is a well-known optical technique \cite{Eichler1986} that has been used with conventional lasers in the visible and IR ranges for over half a century. Following the advent of free-electron lasers operating in the EUV and x-ray ranges, this technique was extended to short wavelengths \cite{Bencivenga2019,Rouxel2021,Li2026}, making it possible to generate transient gratings with nanoscale periodicity. While the initial focus of the short-wavelength TG development was on the generation of high-wave-vector coherent phonons and nanoscale thermal transport \cite{Bencivenga2019,Foglia2023}, it was soon demonstrated that EUV TG excitation of magnetic samples leads to spatially periodic demagnetization, thus creating nanoscale transient magnetization gratings \cite{Ksenzov2021,Yao2022}. It had been previously shown \cite{vanKampen2002} that laser-induced ultrafast demagnetization changes the magnetic anisotropy, which can be exploited to drive magnetic precession. In a TG geometry, launching precession at the maxima of the periodic excitation pattern results in counter-propagating magnons with wavelength \(L\) \cite{Cao2021}.

The excitation of coherent nanoscale magnons in ferrimagnetic Fe/Gd multilayers in an EUV TG experiment has been demonstrated in a recent study \cite{Miedaner2024} conducted at the FERMI free-electron laser facility. The probe wavelength \(\lambda_{\mathrm{pr}} = 8.34\)\,nm was tuned to the N\textsubscript{4,5} absorption edge of Gd, which yields a large magneto-optical effect \cite{Prieto2002}. The probing process, i.e. resonant inelastic EUV/x-ray scattering by magnons, is essentially the same as in the magnon momentum microscopy technique described in the preceding section~\ref{sec:mmm}. As in the latter method, the probe beam is scattered into a well-defined point on the detector controlled by the magnon wave vector. In contrast to MMM, however, the magnon wave vector \(q = 2\pi/L\) is known from the outset; hence, measuring the wave vector is not the objective. Rather, the temporal dynamics of the magnetization are monitored by measuring the intensity of the scattered probe pulse as a function of the pump--probe delay. In the resulting waveforms shown in figures~\ref{fig:euvtg}(b) and (c), one can directly see magnon oscillations. The measurements were done at different wave vectors, with the wave-vector control accomplished by varying the EUV pump wavelength between 25 and 42\,nm. Figure~\ref{fig:euvtg}(d) shows the measured magnon frequency vs. wave vector, i.e. the magnon dispersion curves, for two samples with slightly different Fe/Gd ratios. (The zero-wave-vector data were obtained in conventional optical pump--probe measurements with time-resolved Faraday-rotation detection \cite{Miedaner2024}.) The nearly parabolic dispersion curves indicate that, in the wave-vector range of the measurements, the exchange interaction was already dominant. Sample~2, with a higher Fe content, was closer to the ferrimagnetic compensation point and, as expected, yielded a higher magnon stiffness \cite{Miedaner2024}.

The time-domain detection of the magnon oscillations avoids the need for a high-resolution spectrometer. In effect, the resolution is limited by the length of the delay line: the 1\,ns pump--probe delay range at FERMI corresponds to a frequency resolution of about 1\,GHz, or 4\,\textmu{}eV, whereas the state-of-the-art resolution of inelastic neutron scattering spectroscopy of magnons is about 1\,meV \cite{Man2017}, and that of RIXS is in the tens of meV range \cite{Ament2011,Schlappa2024, Martinelli2022}. The highest magnon wave vector achieved in the initial demonstration \cite{Miedaner2024} was 120\,rad\,\textmu{}m$^{-1}$, corresponding to a wavelength of 52\,nm.

As an `all-optical' technique, EUV TG magnon spectroscopy occupies a special place among the coherent magnon spectroscopy techniques described in this article, as it does not require the fabrication of nanoscale structures such as microwave antennas for launching coherent magnons or Mie resonator structures. As such, it is more suitable for basic physics and materials science studies than for the characterization of magnonic devices.

\subsection{Limitations}\label{euvtg:limitations}

As mentioned above, the technique's major advantage, i.e. the combination of EUV excitation of coherent magnons with EUV probing, is also its major limitation, making it incompatible with microwave-based magnon devices. A significant limitation of EUV TG measurements in the transmission geometry described in Ref.~\cite{Miedaner2024} is that, owing to the strong absorption of EUV radiation in any solid material, the sample must be in the form of a thin membrane. (In Ref.~\cite{Miedaner2024}, 20\,nm thick Fe/Gd multilayers were deposited on 50\,nm thick silicon nitride membranes.) This constraint severely limits the range of samples but can be overcome by using a reflection geometry. The accessible wave-vector range is controlled by the pump and probe wavelengths: the hard limit is given by the smaller of \(2k_{\mathrm{pump}}\) and \(2k_{\mathrm{probe}}\), where \(k_{\mathrm{pump}}\) and \(k_{\mathrm{probe}}\) are the pump and probe wave vectors, respectively. In practice, the smallest EUV TG period demonstrated at FERMI is 17.5\,nm \cite{Miedaner2024}, corresponding to \(q = 360\)\,rad\,\textmu{}m$^{-1}$, but magnon oscillations at this wave vector have not yet been observed, possibly because of overdamping of high-frequency magnons in the Fe/Gd samples. Finally, a major limitation is the need for a free-electron laser facility producing high-intensity, ultrashort, coherent EUV pulses. While there has been a lot of progress in table-top high-harmonic-generation EUV sources \cite{Johnsen2023}, table-top EUV TG measurements have not yet been demonstrated.

\subsection{Perspective}\label{euvtg:perspective}

EUV TG spectroscopy of nanoscale coherent magnons is still in its infancy: as of now, only the initial proof-of-principle study has been published. Yet several avenues for further development can already be anticipated. Firstly, the extension of the TG approach into the soft and hard x-ray ranges is already under way. A very recent study \cite{Miedaner2026} demonstrated hard-x-ray TG excitation of coherent magnons in a garnet film; this experiment used an 800\,nm optical probe pulse; therefore, the magnon wavelength was relatively large (1.3\,\textmu{}m). Interestingly, the magnon generation mechanism was totally different from that in Ref.~\cite{Miedaner2024}: the x-ray excitation launched strain waves (coherent phonons), which drove magnetic precession via magnetostriction. In a concurrent development, all-hard-x-ray TG measurements of coherent phonons with wavelengths of \(\sim\)10\,nm have just been demonstrated \cite{Li2026}. Thus, it is likely that x-ray TG spectroscopy of coherent magnons with wavelengths in the single-digit nanometre range will be possible in the near future. The capabilities of EUV (and eventually soft-x-ray) TG magnon spectroscopy will be greatly enhanced by implementing a reflection geometry that will enable measurements on bulk samples or thin films grown on bulk substrates. EUV TG measurements of coherent surface phonons in a reflection geometry have already been demonstrated \cite{Maznev2021}. While the latter study was done with a near-normal incidence of the probe beam, a grazing-incidence geometry may be beneficial, especially in the soft-x-ray range, as grazing incidence will both greatly increase reflectivity and relax the constraints on the surface roughness of the sample. Finally, it would be exciting to use the TG technique to explore nonlinear regimes and magnon--magnon interactions mentioned in the magnon momentum microscopy section.

\section{Scanning transmission x-ray microscopy}\label{sec:stxm}

\begin{figure}[htbp]
\centering
\includegraphics[width=\textwidth]{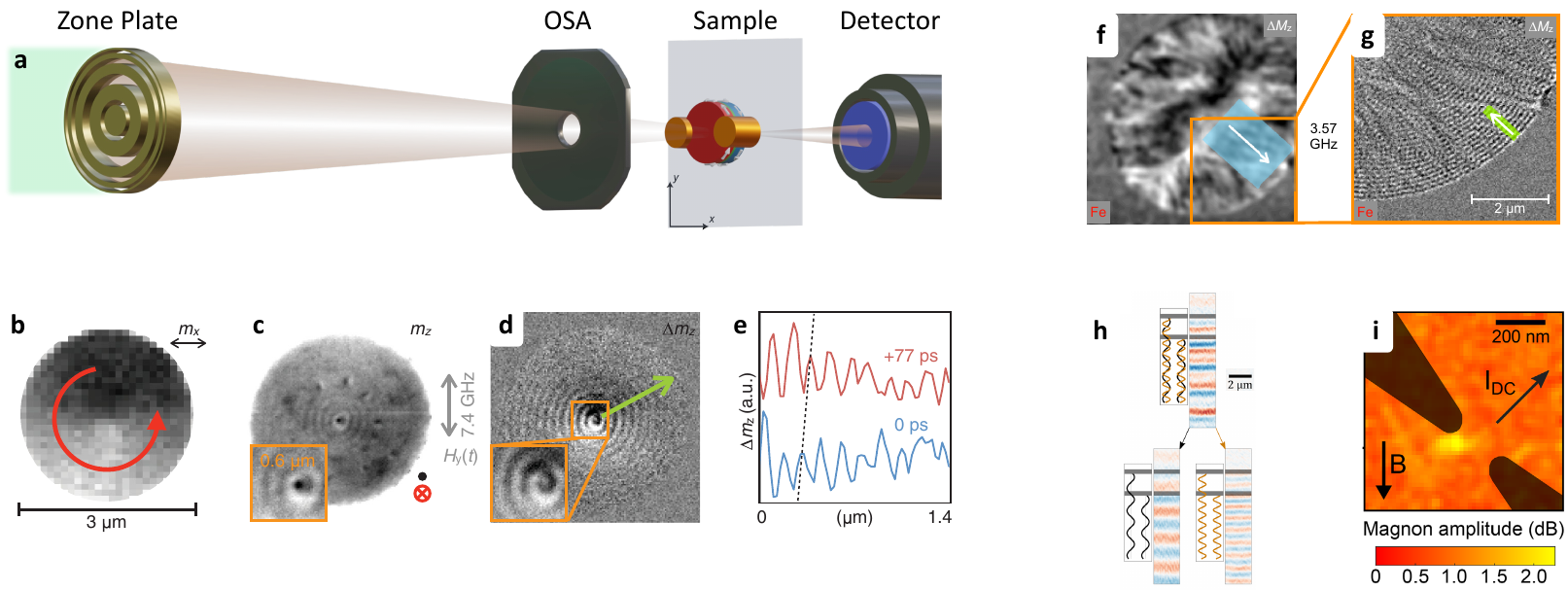}
\caption{\textbf{STXM imaging of coherent magnons.} (a) STXM schematic setup. (b)--(e) Ultrashort heterosymmetric spin waves in a permalloy vortex structure (adapted from \cite{Dieterle2019}). Copyright (2019) by the American Physical Society. (b) Static magnetic vortex configuration, (c) magnons in absorption contrast at 7.4\,GHz and (d) normalized dynamic magnetization contrast. (e) Line profiles at different delay times. (f), (g) Nonreciprocal magnons in a synthetic ferrimagnet (adapted from \cite{Gallardo2024} CC BY 4.0). (f) Outward-propagating long-wavelength magnons, (g) inward-propagating short-wavelength magnons. (h) Analogue magnon adder with frequency-division multiplexing (2.72 and 3.07\,GHz) (adapted from \cite{Schulz2023aip} CC BY 4.0). (i) Phase-locked magnon dynamics at 6.27\,GHz in a direct-current-driven spin Hall nano-oscillator (adapted from \cite{Gonzalez2026} CC BY 4.0).}
\label{fig:stxm}
\end{figure}

Scanning transmission x-ray microscopy (STXM) is well suited for the time-resolved imaging of reproducible magnetization dynamics, offering a unique combination of spatial and temporal resolution (\(\sim\)25\,nm and \(\sim\)10\,ps) \cite{Weigand2022,Finizio2026}. While STXM can in principle be implemented with laboratory-based x-ray sources, it is typically operated at electron storage rings (synchrotrons), which to date provide a much higher brilliance. For magnonics, the focus lies on soft x-rays, as the magnetic contrast is generally higher than for hard x-rays and focusing is easier to achieve. The two key elements of a beamline delivering x-rays to the microscope are the source, which can be a bending magnet or, preferably, an undulator insertion device, and the monochromator, which selects the desired photon energy. Owing to the short attenuation length of soft x-rays in air, the microscope itself is housed in a vacuum vessel and consists of a diffractive focusing lens (Fresnel zone plate), an order-selecting aperture that blocks undiffracted x-rays and higher diffraction orders, and a point transmission detector (see figure~\ref{fig:stxm}(a)). The sample is raster-scanned through the focal spot under interferometric position control, and the locally transmitted intensities yield a two-dimensional image. In addition to absorption contrast, which depends on thickness and atomic number, STXM provides elemental and even chemical selectivity by tuning the photon energy to the resonant absorption edges of the elements of interest. With control of the x-ray polarization, both linear and circular dichroism can be exploited, i.e. the difference in absorption between different polarization states. In this way, the local magnetization of a sample can be imaged via x-ray magnetic circular dichroism (XMCD), even quantitatively in terms of spin and orbital moments \cite{Schutz1987}. XMCD further allows for the sublattice-resolved detection of ferrimagnets and for resolving time-reversal symmetry breaking in altermagnets and noncollinear antiferromagnets \cite{Yamamoto2025,Yamamoto2026,Birch2026}, whereas x-ray magnetic linear dichroism (XMLD) enables imaging of the local spin axis (Néel vector) in antiferromagnets \cite{vanderLaan1986}. During STXM imaging, the sample can be subjected to magnetic and electric fields, electric currents, and high or low temperatures. Moreover, STXM allows for the imaging of buried layers and, using environmental cells, of samples in gas or liquid environments. Related x-ray microscopy techniques include full-field transmission x-ray microscopy, three-dimensional approaches (tomography and laminography) \cite{Koraltan2025,Xu2026}, holography, ptychography, and x-ray photoemission electron microscopy (XPEEM).

For time-resolved (TR) imaging, TR-STXM exploits the inherent time structure of the electron bunches in the storage ring for stroboscopic pump--probe detection \cite{VanWaeyenberge2006,Acremann2006,Kammerer2011,Cheng2012,Bonetti2015}. The duration of the resulting x-ray flashes is typically of the order of 100\,ps and, in dedicated low-alpha operation modes (for example at BESSY II), can be reduced to about 10\,ps \cite{Mayr2024}, at repetition frequencies of typically 500\,MHz. Using fast avalanche photodiodes, the individual x-ray flashes can be detected and sorted according to their delay with respect to the excitation; however, reaching a sufficient signal-to-noise ratio requires integration over many repetitions of the process. In this way, the magnetization (or Néel vector) can be imaged phase-resolved with a unique combination of spatial and temporal resolution, in the bulk of the sample and with elemental and sublattice sensitivity. The related techniques mentioned above can, in principle, also be operated in a TR fashion \cite{Stoll2004,Choe2004,Buttner2015,Donnelly2020,Butcher2025}.

In the field of magnonics, TR-STXM has been used to investigate nanoscale magnetization dynamics that are at present inaccessible to other techniques. This concerns in particular the phase-resolved imaging of coherent magnons with wavelengths below the optical resolution limit \cite{Mayr2024,Wintz2016,Sluka2019,Dieterle2019,Forster2019prb,Baumgaertl2020,Mayr2021,Gallardo2024,Koraltan2024}. Selected observations include the excitation of magnons with nanoscale wavelengths from magnetic vortex cores and domain walls \cite{Mayr2024,Wintz2016,Sluka2019,Dieterle2019,Forster2019prb,Mayr2021,Behncke2018,Albisetti2018}, the propagation of magnons in one-dimensional domain-wall conduits \cite{Sluka2019,Albisetti2018,Che2025}, coherent magnons with ultrashort wavelengths down to 70\,nm \cite{Mayr2024}, strongly nonreciprocal magnon propagation in heterostructures \cite{Mayr2024,Wintz2016,Sluka2019,Mayr2021,Gallardo2024,Koraltan2024,Albisetti2020,Talapatra2023}, the quantitative determination of spin-wave precession angles \cite{Gross2019}, magnon dynamics in single-crystalline insulators \cite{Forster2019prb,Baumgaertl2020,Forster2019jap,Schulz2023aip,Greil2025,Wojewoda2026} and two-dimensional materials \cite{Schulz2023nl}, dynamics in spin-torque \cite{Bonetti2015} and spin Hall nano-oscillators \cite{Gonzalez2026}, magnons in nanotubes \cite{Korber2021}, three-dimensional helices \cite{MoralesFernandez2026}, confined geometries \cite{Trager2020pssr,Trager2021,Pile2022}, and biological systems \cite{Feggeler2023}. Furthermore, magnonic logic devices have been imaged in operando \cite{Schulz2023aip,Talmelli2020}, as have magnonic (quasi)crystals \cite{Lisiecki2019} and the diffractive focusing of spin waves \cite{Grafe2020}.

\subsection{Limitations}\label{stxm:limitations}

A severe limitation of STXM is of a practical nature: since the method relies on synchrotron storage rings, where access is granted through competitive beamtime proposals, the available experimental time is scarce. A second practical constraint is the need for thin, soft-x-ray-transparent samples. This is readily met by polycrystalline films or multilayers deposited on SiN membranes, but is more demanding for single-crystalline materials, which require focused-ion-beam thinning of transparent windows or the extraction of lamellae. On the technical side, although TR-STXM offers a unique combination of spatial and temporal resolution, both still impose significant restrictions on magnon detection, in particular the temporal resolution of \(\sim\)10\,ps. A further issue is the limited signal-to-noise ratio, which precludes single-shot detection and requires integration over many repetitions of the process. As a consequence, TR-STXM is only sensitive to coherent, phase-stable spin waves and is blind to incoherent magnon excitations.

\subsection{Perspective}\label{stxm:perspective}

The spatial resolution of purely absorption-based STXM, which is set by the zone plate, can be overcome by ptychographic detection. Here, the diffracted intensity is additionally recorded with a two-dimensional detector for overlapping, defocused illumination spots, and a computational reconstruction yields amplitude and phase images whose resolution is ultimately limited only by the x-ray wavelength. Ptychography can further be combined with three-dimensional imaging (ptycho-tomography). With present detector technology, however, the accessible excitation frequencies in this extended imaging mode are restricted to multiples of the 500\,MHz synchrotron base frequency. If time resolution is not required, on the other hand, ptycho-tomography could be combined with the recently introduced magnon momentum microscopy (MMM; see section~\ref{sec:mmm}) to achieve spatially resolved detection. The temporal resolution of conventional STXM might be improved beyond the x-ray flash duration by time-tagging the arrival of individual photons. In parallel, two-dimensional detectors are being developed with the aim of resolving the 500\,MHz probe frequency and, eventually, of tagging sub-flash arrival times in each pixel. In the medium to long term, x-ray imaging of magnons could thus approach resolutions of the order of 1\,nm in space and 1\,ps in time. Sensitivity to incoherent magnons might be achieved through correlated noise spectroscopy. New synchrotron sources with dedicated timing modes, such as the planned BESSY III facility in Berlin, high-repetition-rate free-electron lasers, and possibly further improved laboratory sources, will support these developments.

\section{Comparison of the techniques}\label{sec:comparison}

Table~\ref{tab:comparison} summarizes the capabilities and limitations of the techniques discussed in this Perspective. The comparison highlights that wave-vector reach alone is not sufficient to define experimental usefulness; each method offers a different balance between spatial and temporal resolution, frequency sensitivity, coherence selectivity, wave-vector information, and experimental practicality.

\begin{table}[htbp]
\caption{ Comparison of the presented techniques in terms of coherence,
wave-vector selectivity, infrastructure requirements, and main advantages and limitations. Entries refer to current demonstrated implementations.
Abbreviations: \textmu{}BLS, microfocused Brillouin light scattering; PR-BLS, phase-resolved BLS; NV, nitrogen-vacancy; $T_1$, $T_2$, sensor spin relaxation and dephasing times; $d$, sensor--sample distance; EUV, extreme ultraviolet; FEL, free-electron laser.}
\label{tab:comparison}
\centering
\scriptsize
\begin{tabularx}{\textwidth}{@{}>{\raggedright\arraybackslash}p{1.6cm}*{5}{>{\raggedright\arraybackslash}X}@{}}
\toprule
Criterion & \textbf{Mie-enhanced \textmu{}BLS} & \textbf{NV/colour-centre magnetometry} & \textbf{EUV transient grating} & \textbf{Magnon momentum microscopy} & \textbf{Scanning transmission x-ray microscopy} \\
\midrule
Coherence & Thermal/incoherent magnons; coherently driven magnons; phase possible with PR-BLS & Coherently driven spin waves; incoherent/thermal magnons via $T_1$ relaxometry and $T_2$ dephasing spectroscopy & Coherent magnons only, generated by the transient grating & Mainly magnon population/intensity in $k$-space; phase measurements would require synchronized excitation and detection, similar to STXM & Coherent only \\
Wave-vector selectivity & Full magnitude and direction with nanostripes; discrete, grating-selected $k$ & $k$-dependent filtering set by the sensor--sample distance, with maximum sensitivity around $k \sim 1/d$ & One selected $k$ per grating period/orientation & Direct 2D momentum-space map & Fourier transform of the real-space image \\
Infrastructure requirements & Table-top \textmu{}BLS and nanofabricated dielectric resonators & Scanning or widefield fluorescence microscope & EUV/FEL facility & Soft-x-ray beamline at a synchrotron, FEL or laser-driven source & Synchrotron \\
Main advantage & High frequency; standard table-top setup & Nanoscale local magnetic-field imaging with high sensitivity & High time resolution & 2D information up to very large wave vectors; no frequency limitation; very simple & High spatial and temporal resolution \\
Main limitation & Needs nanofabrication of dielectric structures & Protocol-dependent frequency range & Facility-intensive; selected $k$ only; EUV-transparent sample & No direct phase/time information in base mode; limited spatial resolution; no frequency information; x-ray-transparent sample & x-ray-transparent sample \\
\bottomrule
\end{tabularx}
\end{table}

\section{Conclusion}\label{sec:conclusion}
The techniques discussed in this Perspective probe complementary properties of spin waves, and together they now reach into the wave-vector range that was previously inaccessible. Several of the discussed techniques now resolve spin waves with wavelengths of a few tens of nanometres, well below the optical diffraction limit, and all of them are steadily progressing towards shorter wavelengths.

The methods differ not only how, but also what they measure. NV/colour-centre magnetometry and STXM image spin waves in real space with a resolution of a few tens of nanometres. NV magnetometry does so through the stray field, including beneath optically opaque layers, while STXM offers element and sublattice selectivity. MMM records the full two-dimensional momentum distribution, and EUV TG follows the magnetization dynamics at a fixed wave vector in the time domain. Mie-enhanced \textmu{}BLS resolves the magnon frequency directly. Together with NV relaxometry and MMM, it is also sensitive to incoherent magnons.

The temporal and spectral reach of the methods is equally diverse. STXM and EUV TG achieve picosecond and sub-picosecond resolution, respectively, whereas MMM and NV magnetometry currently operate under steady-state excitation. NV and STXM measurements are so far confined to frequencies of a few to tens of gigahertz. Mie-enhanced \textmu{}BLS and EUV TG access considerably higher frequencies, and MMM has no intrinsic frequency limit.

Choosing a technique is therefore governed less by wave-vector reach than by the question at hand. Antenna-based approaches (Mie-enhanced \textmu{}BLS, NV magnetometry, MMM and STXM) are suited to characterizing magnonic devices in operation. The all-optical EUV TG approach instead probes intrinsic material properties without any fabricated excitation structures.

On the other hand, improved excitation sources of coherent short-wavelength spin waves, such as grating couplers, spin textures and parametric pumping, will further allow the techniques that can measure such excitations to be benchmarked, revealing their true potential and accessible range.

As the described techniques mature from proof-of-principle demonstrations into routine experimental tools, they will open the exchange-dominated regime to systematic study. This includes nonlinear magnon interactions, chiral magnons in altermagnets, and high-frequency antiferromagnetic and van der Waals magnets. They will also provide the metrology needed to scale magnon-based computing down to genuinely nanoscale wavelengths.

\funding{O.W. was supported by Horizon Europe -- MSCA grant agreement No.~101211677, project FeriMag. S. K. acknowledges support from the Dutch Research Council (NWO) via VI.Veni.222.296.}
\data{No new data were created}

\bibliographystyle{iopart-num}
\bibliography{references}

@article{Chumak2015,
  author  = {Chumak, A. V. and Vasyuchka, V. I. and Serga, A. A. and Hillebrands, B.},
  title   = {{Magnon spintronics}},
  journal = {Nature Physics},
  volume  = {11},
  pages   = {453--461},
  year    = {2015},
  doi     = {10.1038/nphys3347}
}

@article{Pirro2021,
  author  = {Pirro, P. and Vasyuchka, V. I. and Serga, A. A. and Hillebrands, B.},
  title   = {{Advances in coherent magnonics}},
  journal = {Nature Reviews Materials},
  volume  = {6},
  pages   = {1114--1135},
  year    = {2021},
  doi     = {10.1038/s41578-021-00332-w}
}

@article{DamonEshbach1961,
  author  = {Damon, R. W. and Eshbach, J. R.},
  title   = {{Magnetostatic modes of a ferromagnet slab}},
  journal = {Journal of Physics and Chemistry of Solids},
  volume  = {19},
  pages   = {308--320},
  year    = {1961},
  doi     = {10.1016/0022-3697(61)90041-5}
}

@article{Kalinikos1986,
  author  = {Kalinikos, B. A. and Slavin, A. N.},
  title   = {{Theory of dipole-exchange spin wave spectrum for ferromagnetic films with mixed exchange boundary conditions}},
  journal = {Journal of Physics C: Solid State Physics},
  volume  = {19},
  pages   = {7013--7033},
  year    = {1986},
  doi     = {10.1088/0022-3719/19/35/014}
}

@article{Serga2010,
  author  = {Serga, A. A. and Chumak, A. V. and Hillebrands, B.},
  title   = {{YIG magnonics}},
  journal = {Journal of Physics D: Applied Physics},
  volume  = {43},
  pages   = {264002},
  year    = {2010},
  doi     = {10.1088/0022-3727/43/26/264002}
}

@article{Sebastian2015,
  author  = {Sebastian, T. and Schultheiss, K. and Obry, B. and Hillebrands, B. and Schultheiss, H.},
  title   = {{Micro-focused Brillouin light scattering: imaging spin waves at the nanoscale}},
  journal = {Frontiers in Physics},
  volume  = {3},
  pages   = {35},
  year    = {2015},
  doi     = {10.3389/fphy.2015.00035}
}

@article{Krcma2025,
  author  = {Kr{\v{c}}ma, J. and Wojewoda, O. and Hrto{\v{n}}, M. and Holobr{\'a}dek, J. and Arregi, J. A. and Panda, J. and Pribytova, E. and Urb{\'a}nek, M.},
  title   = {{Mie-enhanced microfocused Brillouin light scattering for full wave vector resolution of nanoscale spin waves}},
  journal = {Science Advances},
  volume  = {11},
  pages   = {eady8833},
  year    = {2025},
  doi     = {10.1126/sciadv.ady8833}
}

@article{Smejkal2023,
  author  = {{\v{S}}mejkal, L. and Marmodoro, A. and Ahn, K.-H. and Gonz{\'a}lez-Hern{\'a}ndez, R. and Turek, I. and Mankovsky, S. and Ebert, H. and D'Souza, S. W. and {\v{S}}ipr, O. and Sinova, J. and Jungwirth, T.},
  title   = {{Chiral Magnons in Altermagnetic RuO$_2$}},
  journal = {Physical Review Letters},
  volume  = {131},
  pages   = {256703},
  year    = {2023},
  doi     = {10.1103/physrevlett.131.256703}
}

@article{Liu2024,
  author  = {Liu, Z. and Ozeki, M. and Asai, S. and Itoh, S. and Masuda, T.},
  title   = {{Chiral Split Magnon in Altermagnetic MnTe}},
  journal = {Physical Review Letters},
  volume  = {133},
  pages   = {156702},
  year    = {2024},
  doi     = {10.1103/physrevlett.133.156702}
}

@article{Wittrock2026,
  author  = {Wittrock, S. and Klose, C. and Perna, S. and Baumgaertl, K. and Mucchietto, A. and Schneider, M. and Fuchs, J. and Deinhart, V. and Karaman, T. and Grundler, D. and Eisebitt, S. and Pfau, B. and Schick, D.},
  title   = {{Soft-X-ray momentum microscopy of nonlinear magnon interactions}},
  journal = {Nature Physics},
  volume  = {22},
  pages   = {1160--1165},
  year    = {2026},
  doi     = {10.1038/s41567-026-03318-z}
}

@article{Bertelli2020,
  author  = {Bertelli, I. and Carmiggelt, J. J. and Yu, T. and Simon, B. G. and Pothoven, C. C. and Bauer, G. E. W. and Blanter, Y. M. and Aarts, J. and van der Sar, T.},
  title   = {{Magnetic resonance imaging of spin-wave transport and interference in a magnetic insulator}},
  journal = {Science Advances},
  volume  = {6},
  pages   = {eabd3556},
  year    = {2020},
  doi     = {10.1126/sciadv.abd3556}
}

@article{ManasValero2026,
  author  = {Ma{\~n}as-Valero, S. and Doedes, Y. C. and Bondarenko, A. and Borst, M. and Kurdi, S. and Poirier, T. and Edgar, J. H. and Jacques, V. and Blanter, Y. M. and van der Sar, T.},
  title   = {{Isofrequency spin-wave imaging using color center magnetometry for magnon spintronics}},
  journal = {Nature Communications},
  volume  = {17},
  pages   = {379},
  year    = {2026},
  doi     = {10.1038/s41467-025-67056-1}
}

@article{SimonKurdi2022,
  author  = {Simon, B. G. and Kurdi, S. and Carmiggelt, J. J. and Borst, M. and Katan, A. J. and van der Sar, T.},
  title   = {{Filtering and Imaging of Frequency-Degenerate Spin Waves Using Nanopositioning of a Single-Spin Sensor}},
  journal = {Nano Letters},
  volume  = {22},
  pages   = {9198--9204},
  year    = {2022},
  doi     = {10.1021/acs.nanolett.2c02791}
}

@article{Miedaner2024,
  author  = {Miedaner, P. R. and Berndt, N. and Deschamps, J. and Urazhdin, S. and Khatu, N. and Fainozzi, D. and Brioschi, M. and Carrara, P. and Cucini, R. and Rossi, G. and Wittrock, S. and Ksenzov, D. and Mincigrucci, R. and Bencivenga, F. and Foglia, L. and Paltanin, E. and Bonetti, S. and Engel, D. and Schick, D. and Gutt, C. and Comin, R. and Nelson, K. A. and Maznev, A. A.},
  title   = {{Excitation and detection of coherent nanoscale spin waves via extreme ultraviolet transient gratings}},
  journal = {Science Advances},
  volume  = {10},
  pages   = {eadp6015},
  year    = {2024},
  doi     = {10.1126/sciadv.adp6015}
}

@article{Sluka2019,
  author  = {Sluka, V. and Schneider, T. and Gallardo, R. A. and K{\'a}kay, A. and Weigand, M. and Warnatz, T. and Mattheis, R. and Rold{\'a}n-Molina, A. and Landeros, P. and Tiberkevich, V. and Slavin, A. and Sch{\"u}tz, G. and Erbe, A. and Deac, A. and Lindner, J. and Raabe, J. and Fassbender, J. and Wintz, S.},
  title   = {{Emission and propagation of 1D and 2D spin waves with nanoscale wavelengths in anisotropic spin textures}},
  journal = {Nature Nanotechnology},
  volume  = {14},
  pages   = {328--333},
  year    = {2019},
  doi     = {10.1038/s41565-019-0383-4}
}

@article{Trager2020,
  author  = {Tr{\"a}ger, N. and Gro{\ss}, F. and F{\"o}rster, J. and Baumgaertl, K. and Stoll, H. and Weigand, M. and Sch{\"u}tz, G. and Grundler, D. and Gr{\"a}fe, J.},
  title   = {{Single shot acquisition of spatially resolved spin wave dispersion relations using X-ray microscopy}},
  journal = {Scientific Reports},
  volume  = {10},
  pages   = {18146},
  year    = {2020},
  doi     = {10.1038/s41598-020-74785-4}
}

@article{Brillouin1922,
  author  = {Brillouin, L.},
  title   = {{Diffusion de la lumi{\`e}re et des rayons X par un corps transparent homog{\`e}ne: Influence de l'agitation thermique}},
  journal = {Annales de Physique},
  volume  = {9},
  pages   = {88--122},
  year    = {1922},
  doi     = {10.1051/anphys/192209170088}
}

@article{Merklein2022,
  author  = {Merklein, M. and Kabakova, I. V. and Zarifi, A. and Eggleton, B. J.},
  title   = {{100 years of Brillouin scattering: Historical and future perspectives}},
  journal = {Applied Physics Reviews},
  volume  = {9},
  pages   = {041306},
  year    = {2022},
  doi     = {10.1063/5.0095488}
}

@article{Lindsay1981,
  author  = {Lindsay, S. M. and Anderson, M. W. and Sandercock, J. R.},
  title   = {{Construction and alignment of a high performance multipass vernier tandem Fabry--Perot interferometer}},
  journal = {Review of Scientific Instruments},
  volume  = {52},
  pages   = {1478--1486},
  year    = {1981},
  doi     = {10.1063/1.1136479}
}

@article{ChagnonLessard2026,
  author  = {Chagnon-Lessard, S. and Nicolai, J. and Steller, J. and Moo, E. K. and Jean-Ruel, H.},
  title   = {{High-performance Brillouin spectroscopy using VIPA-etalon cascades}},
  journal = {Journal of Biomedical Optics},
  volume  = {31},
  pages   = {026001},
  year    = {2026},
  doi     = {10.1117/1.jbo.31.2.026001}
}

@article{Wojewoda2023,
  author  = {Wojewoda, O. and Ligmajer, F. and Hrto{\v{n}}, M. and Kl{\'\i}ma, J. and Dhankhar, M. and Dav{\'\i}dkov{\'a}, K. and Sta{\v{n}}o, M. and Holobr{\'a}dek, J. and Kr{\v{c}}ma, J. and Zl{\'a}mal, J. and {\v{S}}ikola, T. and Urb{\'a}nek, M.},
  title   = {{Observing high-k magnons with Mie-resonance-enhanced Brillouin light scattering}},
  journal = {Communications Physics},
  volume  = {6},
  pages   = {94},
  year    = {2023},
  doi     = {10.1038/s42005-023-01214-z}
}

@article{Wojewoda2024Modeling,
  author  = {Wojewoda, O. and Hrto{\v{n}}, M. and Urb{\'a}nek, M.},
  title   = {{Modeling of microfocused Brillouin light scattering spectra}},
  journal = {Physical Review B},
  volume  = {110},
  pages   = {224428},
  year    = {2024},
  doi     = {10.1103/physrevb.110.224428}
}

@article{Madami2012,
  author  = {Madami, M. and Gubbiotti, G. and Tacchi, S. and Carlotti, G.},
  title   = {{Application of Microfocused Brillouin Light Scattering to the Study of Spin Waves in Low-Dimensional Magnetic Systems}},
  journal = {Solid State Physics},
  volume  = {63},
  pages   = {79--150},
  year    = {2012},
  doi     = {10.1016/b978-0-12-397028-2.00002-3}
}

@article{Wojewoda2023phase,
  author  = {Wojewoda, O. and Hrto{\v{n}}, M. and Dhankhar, M. and Kr{\v{c}}ma, J. and Dav{\'\i}dkov{\'a}, K. and Kl{\'\i}ma, J. and Holobr{\'a}dek, J. and Ligmajer, F. and {\v{S}}ikola, T. and Urb{\'a}nek, M.},
  title   = {{Phase-resolved optical characterization of nanoscale spin waves}},
  journal = {Applied Physics Letters},
  volume  = {122},
  pages   = {202405},
  year    = {2023},
  doi     = {10.1063/5.0151338}
}

@article{Freeman2020,
  author  = {Freeman, R. and Lemasters, R. and Kalejaiye, T. and Wang, F. and Chen, G. and Ding, J. and Wu, M. and Demidov, V. E. and Demokritov, S. O. and Harutyunyan, H. and Urazhdin, S.},
  title   = {{Brillouin light scattering of spin waves inaccessible with free-space light}},
  journal = {Physical Review Research},
  volume  = {2},
  pages   = {033427},
  year    = {2020},
  doi     = {10.1103/physrevresearch.2.033427}
}

@article{Lozovski2024,
  author  = {Lozovski, V. and Chumak, A. V.},
  title   = {{Plasmon-enhanced Brillouin light scattering spectroscopy for magnetic systems: Theoretical model}},
  journal = {Physical Review B},
  volume  = {110},
  pages   = {184419},
  year    = {2024},
  doi     = {10.1103/physrevb.110.184419}
}

@article{Jersch2010,
  author  = {Jersch, J. and Demidov, V. E. and Fuchs, H. and Rott, K. and Krzysteczko, P. and M{\"u}nchenberger, J. and Reiss, G. and Demokritov, S. O.},
  title   = {{Mapping of localized spin-wave excitations by near-field Brillouin light scattering}},
  journal = {Applied Physics Letters},
  volume  = {97},
  pages   = {152502},
  year    = {2010},
  doi     = {10.1063/1.3502599}
}

@article{Jeanmaire1977,
  author  = {Jeanmaire, D. L. and Van Duyne, R. P.},
  title   = {{Surface Raman spectroelectrochemistry: Part I. Heterocyclic, aromatic, and aliphatic amines adsorbed on the anodized silver electrode}},
  journal = {Journal of Electroanalytical Chemistry and Interfacial Electrochemistry},
  volume  = {84},
  pages   = {1--20},
  year    = {1977},
  doi     = {10.1016/s0022-0728(77)80224-6}
}

@article{LeRu2007,
  author  = {Le Ru, E. C. and Blackie, E. and Meyer, M. and Etchegoin, P. G.},
  title   = {{Surface Enhanced Raman Scattering Enhancement Factors:~ A Comprehensive Study}},
  journal = {The Journal of Physical Chemistry C},
  volume  = {111},
  pages   = {13794--13803},
  year    = {2007},
  doi     = {10.1021/jp0687908}
}

@article{Hollander2018,
  author  = {Holl{\"a}nder, R. B. and M{\"u}ller, C. and Schmalz, J. and Gerken, M. and McCord, J.},
  title   = {{Magnetic domain walls as broadband spin wave and elastic magnetisation wave emitters}},
  journal = {Scientific Reports},
  volume  = {8},
  pages   = {13871},
  year    = {2018},
  doi     = {10.1038/s41598-018-31689-8}
}

@article{Yu2013,
  author  = {Yu, H. and Duerr, G. and Huber, R. and Bahr, M. and Schwarze, T. and Brandl, F. and Grundler, D.},
  title   = {{Omnidirectional spin-wave nanograting coupler}},
  journal = {Nature Communications},
  volume  = {4},
  pages   = {2702},
  year    = {2013},
  doi     = {10.1038/ncomms3702}
}

@article{Heinz2022,
  author  = {Heinz, B. and Mohseni, M. and Lentfert, A. and Verba, R. and Schneider, M. and L{\"a}gel, B. and Levchenko, K. and Br{\"a}cher, T. and Dubs, C. and Chumak, A. V. and Pirro, P.},
  title   = {{Parametric generation of spin waves in nanoscaled magnonic conduits}},
  journal = {Physical Review B},
  volume  = {105},
  pages   = {144424},
  year    = {2022},
  doi     = {10.1103/physrevb.105.144424}
}

@article{Verma2017,
  author  = {Verma, P.},
  title   = {{Tip-Enhanced Raman Spectroscopy: Technique and Recent Advances}},
  journal = {Chemical Reviews},
  volume  = {117},
  pages   = {6447--6466},
  year    = {2017},
  doi     = {10.1021/acs.chemrev.6b00821}
}

@article{Casola2018,
  author  = {Casola, F. and van der Sar, T. and Yacoby, A.},
  title   = {{Probing condensed matter physics with magnetometry based on nitrogen-vacancy centres in diamond}},
  journal = {Nature Reviews Materials},
  volume  = {3},
  pages   = {17088},
  year    = {2018},
  doi     = {10.1038/natrevmats.2017.88}
}

@article{Zhou2021,
  author  = {Zhou, T. X. and Carmiggelt, J. J. and G{\"a}chter, L. M. and Esterlis, I. and Sels, D. and St{\"o}hr, R. J. and Du, C. and Fernandez, D. and Rodriguez-Nieva, J. F. and B{\"u}ttner, F. and Demler, E. and Yacoby, A.},
  title   = {{A magnon scattering platform}},
  journal = {Proceedings of the National Academy of Sciences},
  volume  = {118},
  pages   = {e2019473118},
  year    = {2021},
  doi     = {10.1073/pnas.2019473118}
}

@article{VanDerSar2015,
  author  = {van der Sar, T. and Casola, F. and Walsworth, R. and Yacoby, A.},
  title   = {{Nanometre-scale probing of spin waves using single electron spins}},
  journal = {Nature Communications},
  volume  = {6},
  pages   = {7886},
  year    = {2015},
  doi     = {10.1038/ncomms8886}
}

@article{Kikuchi2017,
  author  = {Kikuchi, D. and Prananto, D. and Hayashi, K. and Laraoui, A. and Mizuochi, N. and Hatano, M. and Saitoh, E. and Kim, Y. and Meriles, C. A. and An, T.},
  title   = {{Long-distance excitation of nitrogen-vacancy centers in diamond via surface spin waves}},
  journal = {Applied Physics Express},
  volume  = {10},
  pages   = {103004},
  year    = {2017},
  doi     = {10.7567/apex.10.103004}
}

@article{Grinolds2013,
  author  = {Grinolds, M. S. and Hong, S. and Maletinsky, P. and Luan, L. and Lukin, M. D. and Walsworth, R. L. and Yacoby, A.},
  title   = {{Nanoscale magnetic imaging of a single electron spin under ambient conditions}},
  journal = {Nature Physics},
  volume  = {9},
  pages   = {215--219},
  year    = {2013},
  doi     = {10.1038/nphys2543}
}

@article{Cuche2009,
  author  = {Cuche, A. and Drezet, A. and Sonnefraud, Y. and Faklaris, O. and Treussart, F. and Roch, J.-F. and Huant, S.},
  title   = {{Near-field optical microscopy with a nanodiamond-based single-photon tip}},
  journal = {Optics Express},
  volume  = {17},
  pages   = {19969},
  year    = {2009},
  doi     = {10.1364/oe.17.019969}
}

@article{Maletinsky2012,
  author  = {Maletinsky, P. and Hong, S. and Grinolds, M. S. and Hausmann, B. and Lukin, M. D. and Walsworth, R. L. and Loncar, M. and Yacoby, A.},
  title   = {{A robust scanning diamond sensor for nanoscale imaging with single nitrogen-vacancy centres}},
  journal = {Nature Nanotechnology},
  volume  = {7},
  pages   = {320--324},
  year    = {2012},
  doi     = {10.1038/nnano.2012.50}
}

@article{Simon2021,
  author  = {Simon, B. G. and Kurdi, S. and La, H. and Bertelli, I. and Carmiggelt, J. J. and Ruf, M. and de Jong, N. and van den Berg, H. and Katan, A. J. and van der Sar, T.},
  title   = {{Directional Excitation of a High-Density Magnon Gas Using Coherently Driven Spin Waves}},
  journal = {Nano Letters},
  volume  = {21},
  pages   = {8213--8219},
  year    = {2021},
  doi     = {10.1021/acs.nanolett.1c02654}
}

@article{Carmiggelt2023,
  author  = {Carmiggelt, J. J. and Bertelli, I. and Mulder, R. W. and Teepe, A. and Elyasi, M. and Simon, B. G. and Bauer, G. E. W. and Blanter, Y. M. and van der Sar, T.},
  title   = {{Broadband microwave detection using electron spins in a hybrid diamond-magnet sensor chip}},
  journal = {Nature Communications},
  volume  = {14},
  pages   = {490},
  year    = {2023},
  doi     = {10.1038/s41467-023-36146-3}
}

@article{Maze2008,
  author  = {Maze, J. R. and Stanwix, P. L. and Hodges, J. S. and Hong, S. and Taylor, J. M. and Cappellaro, P. and Jiang, L. and Dutt, M. V. G. and Togan, E. and Zibrov, A. S. and Yacoby, A. and Walsworth, R. L. and Lukin, M. D.},
  title   = {{Nanoscale magnetic sensing with an individual electronic spin in diamond}},
  journal = {Nature},
  volume  = {455},
  pages   = {644--647},
  year    = {2008},
  doi     = {10.1038/nature07279}
}

@article{Wolfe2016,
  author  = {Wolfe, C. S. and Manuilov, S. A. and Purser, C. M. and Teeling-Smith, R. and Dubs, C. and Hammel, P. C. and Bhallamudi, V. P.},
  title   = {{Spatially resolved detection of complex ferromagnetic dynamics using optically detected nitrogen-vacancy spins}},
  journal = {Applied Physics Letters},
  volume  = {108},
  pages   = {232409},
  year    = {2016},
  doi     = {10.1063/1.4953108}
}

@article{Du2017,
  author  = {Du, C. and van der Sar, T. and Zhou, T. X. and Upadhyaya, P. and Casola, F. and Zhang, H. and Onbasli, M. C. and Ross, C. A. and Walsworth, R. L. and Tserkovnyak, Y. and Yacoby, A.},
  title   = {{Control and local measurement of the spin chemical potential in a magnetic insulator}},
  journal = {Science},
  volume  = {357},
  pages   = {195--198},
  year    = {2017},
  doi     = {10.1126/science.aak9611}
}

@article{Melendez2025,
  author  = {Melendez, A. L. and Das, S. and Rodriguez, F. A. and Kao, I.-H. and Liu, W. and Williams, A. J. and Lv, B. and Goldberger, J. and Chatterjee, S. and Singh, S. and Hammel, P. C.},
  title   = {{Quantum sensing of broadband spin dynamics and magnon transport in antiferromagnets}},
  journal = {Science Advances},
  volume  = {11},
  pages   = {eadu9381},
  year    = {2025},
  doi     = {10.1126/sciadv.adu9381}
}

@misc{Zhou2026,
  author  = {Zhou, L.-J. and Nair, J. M. P. and Li, S. and Poirier, T. and Zhao, Y. and Xiong, Z. and Rathi, S. and Zhou, J. and Jiang, Z. and Wang, H. and Edgar, J. H. and Flebus, B. and Du, C. R.},
  title   = {{Observation and Control of Spontaneous Magnon Emission from Spin Ensembles in 2D Hexagonal Boron Nitride}},
  howpublished = {arXiv:2606.20996},
  year    = {2026},
  doi     = {10.48550/arXiv.2606.20996}
}

@article{GrubisicCabo2025,
  author  = {Grubi{\v{s}}i{\'c}-{\v{C}}abo, A. and Guimar{\~a}es, M. H. D. and Afanasiev, D. and Garcia Aguilar, J. H. and Aguilera, I. and Ali, M. N. and Bhattacharyya, S. and Blanter, Y. M. and Bosma, R. and Cheng, Z. and Dan, Z. and Dash, S. P. and Medina Due{\~n}as, J. and Fernandez-Rossier, J. and Gibertini, M. and Grytsiuk, S. and Houmes, M. J. A. and Isaeva, A. and Knekna, C. and Kole, A. H. and Kurdi, S. and Lado, J. L. and Ma{\~n}as-Valero, S. and Lopes, J. M. J. and Marian, D. and Na, M. and Pabst, F. and Barquero Pierantoni, S. and Regout, M. and Reho, R. and R{\"o}sner, M. and Sanz, D. and van der Sar, T. and S{\l}awi{\'n}ska, J. and Verstraete, M. J. and Waseem, M. and van der Zant, H. S. J. and Zanolli, Z. and Soriano, D.},
  title   = {{Roadmap on quantum magnetic materials}},
  journal = {2D Materials},
  volume  = {12},
  pages   = {031501},
  year    = {2025},
  doi     = {10.1088/2053-1583/adbe89}
}

@article{Xu2025,
  author  = {Xu, Z. and Palm, M. L. and Huxter, W. and Herb, K. and Abendroth, J. M. and Bouzehouane, K. and Boulle, O. and Gabor, M. S. and Urrestarazu Larranaga, J. and Morales, A. and Rhensius, J. and Puebla-Hellmann, G. and Degen, C. L.},
  title   = {{Minimizing Sensor-Sample Distances in Scanning Nitrogen-Vacancy Magnetometry}},
  journal = {ACS Nano},
  volume  = {19},
  pages   = {8255--8265},
  year    = {2025},
  doi     = {10.1021/acsnano.4c18460}
}

@article{Vogl2026,
  author  = {Vogl, T. and Iv{\'a}dy, V. and Luxmoore, I. J. and Stern, H. L.},
  title   = {{Defects in hexagonal boron nitride for quantum technologies: a perspective}},
  journal = {2D Materials},
  volume  = {13},
  pages   = {023001},
  year    = {2026},
  doi     = {10.1088/2053-1583/ae566e}
}

@article{Gilardoni2025,
  author  = {Gilardoni, C. M. and Eizagirre Barker, S. and Curtin, C. L. and Fraser, S. A. and Powell, O. F. J. and Lewis, D. K. and Deng, X. and Ramsay, A. J. and Adhikari, S. and Li, C. and Aharonovich, I. and Tan, H. H. and Atat{\"u}re, M. and Stern, H. L.},
  title   = {{A single spin in hexagonal boron nitride for vectorial quantum magnetometry}},
  journal = {Nature Communications},
  volume  = {16},
  pages   = {4947},
  year    = {2025},
  doi     = {10.1038/s41467-025-59642-0}
}

@article{Herb2025,
  author  = {Herb, K. and V{\"o}lker, L. A. and Abendroth, J. M. and Meinhardt, N. and van Schie, L. and Gambardella, P. and Degen, C. L.},
  title   = {{Quantum magnetometry of transient signals with a time resolution of 1.1 nanoseconds}},
  journal = {Nature Communications},
  volume  = {16},
  pages   = {822},
  year    = {2025},
  doi     = {10.1038/s41467-025-55956-1}
}

@article{Qin2021,
  author  = {Qin, H. and Holl{\"a}nder, R. B. and Flaj{\v{s}}man, L. and Hermann, F. and Dreyer, R. and Woltersdorf, G. and van Dijken, S.},
  title   = {{Nanoscale magnonic Fabry-P{\'e}rot resonator for low-loss spin-wave manipulation}},
  journal = {Nature Communications},
  volume  = {12},
  pages   = {2293},
  year    = {2021},
  doi     = {10.1038/s41467-021-22520-6}
}

@article{Gross2020,
  author  = {Gro{\ss}, F. and Zelent, M. and Tr{\"a}ger, N. and F{\"o}rster, J. and Sanli, U. T. and Sauter, R. and Decker, M. and Back, C. H. and Weigand, M. and Keskinbora, K. and Sch{\"u}tz, G. and Krawczyk, M. and Gr{\"a}fe, J.},
  title   = {{Building Blocks for Magnon Optics: Emission and Conversion of Short Spin Waves}},
  journal = {ACS Nano},
  volume  = {14},
  pages   = {17184--17193},
  year    = {2020},
  doi     = {10.1021/acsnano.0c07076}
}

@article{Borst2023,
  author  = {Borst, M. and Vree, P. H. and Lowther, A. and Teepe, A. and Kurdi, S. and Bertelli, I. and Simon, B. G. and Blanter, Y. M. and van der Sar, T.},
  title   = {{Observation and control of hybrid spin-wave--Meissner-current transport modes}},
  journal = {Science},
  volume  = {382},
  pages   = {430--434},
  year    = {2023},
  doi     = {10.1126/science.adj7576}
}

@article{Scheidegger2022,
  author  = {Scheidegger, P. J. and Diesch, S. and Palm, M. L. and Degen, C. L.},
  title   = {{Scanning nitrogen-vacancy magnetometry down to 350~mK}},
  journal = {Applied Physics Letters},
  volume  = {120},
  pages   = {224001},
  year    = {2022},
  doi     = {10.1063/5.0093548}
}

@article{Li2024,
  author  = {Li, Y. and Welker, G. and Norte, R. and van der Sar, T.},
  title   = {{A robust, fiber-coupled scanning probe magnetometer using electron spins at the tip of a diamond nanobeam}},
  journal = {New Journal of Physics},
  volume  = {26},
  pages   = {103031},
  year    = {2024},
  doi     = {10.1088/1367-2630/ad8779}
}

@article{Li2023,
  author  = {Li, Y. and Gerritsma, F. A. and Kurdi, S. and Codreanu, N. and Gr{\"o}blacher, S. and Hanson, R. and Norte, R. and van der Sar, T.},
  title   = {{A Fiber-Coupled Scanning Magnetometer with Nitrogen-Vacancy Spins in a Diamond Nanobeam}},
  journal = {ACS Photonics},
  volume  = {10},
  pages   = {1859--1865},
  year    = {2023},
  doi     = {10.1021/acsphotonics.3c00259}
}

@article{Bae2022,
  author  = {Bae, Y. J. and Wang, J. and Scheie, A. and Xu, J. and Chica, D. G. and Diederich, G. M. and Cenker, J. and Ziebel, M. E. and Bai, Y. and Ren, H. and Dean, C. R. and Delor, M. and Xu, X. and Roy, X. and Kent, A. D. and Zhu, X.},
  title   = {{Exciton-coupled coherent magnons in a 2D semiconductor}},
  journal = {Nature},
  volume  = {609},
  pages   = {282--286},
  year    = {2022},
  doi     = {10.1038/s41586-022-05024-1}
}

@article{Mayr2024,
  author  = {Mayr, S. and F{\"o}rster, J. and Finizio, S. and Schultheiss, K. and Gallardo, R. A. and Narkovicz, R. and Dieterle, G. and Semisalova, A. and Bailey, J. and Kirk, E. and Suszka, A. and Lindner, J. and Gr{\"a}fe, J. and Raabe, J. and Sch{\"u}tz, G. and Weigand, M. and Stoll, H. and Wintz, S.},
  title   = {{Time-resolved x-ray imaging of nanoscale spin-wave dynamics at multi-GHz frequencies using low-alpha synchrotron operation}},
  journal = {Applied Physics Reviews},
  volume  = {11},
  pages   = {041411},
  year    = {2024},
  doi     = {10.1063/5.0206576}
}

@article{Biesner2022,
  author  = {Biesner, T. and Roh, S. and Razpopov, A. and Willwater, J. and S{\"u}llow, S. and Li, Y. and Zoch, K. M. and Medarde, M. and Nuss, J. and Gorbunov, D. and Skourski, Y. and Pustogow, A. and Brown, S. E. and Krellner, C. and Valent{\'\i}, R. and Puphal, P. and Dressel, M.},
  title   = {{Multi-Center Magnon Excitations Open the Entire Brillouin Zone to Terahertz Magnetometry of Quantum Magnets}},
  journal = {Advanced Quantum Technologies},
  volume  = {5},
  pages   = {2200023},
  year    = {2022},
  doi     = {10.1002/qute.202200023}
}

@article{Ament2011,
  author  = {Ament, L. J. P. and van Veenendaal, M. and Devereaux, T. P. and Hill, J. P. and van den Brink, J.},
  title   = {{Resonant inelastic x-ray scattering studies of elementary excitations}},
  journal = {Reviews of Modern Physics},
  volume  = {83},
  pages   = {705--767},
  year    = {2011},
  doi     = {10.1103/RevModPhys.83.705}
}

@article{Vollmer2004,
  author  = {Vollmer, R. and Etzkorn, M. and Anil Kumar, P. S. and Ibach, H. and Kirschner, J.},
  title   = {{Spin-polarized electron energy loss spectroscopy: a method to measure magnon energies}},
  journal = {Journal of Magnetism and Magnetic Materials},
  volume  = {272-276},
  pages   = {2126--2130},
  year    = {2004},
  doi     = {10.1016/j.jmmm.2003.12.506}
}

@article{Hashimoto2017,
  author  = {Hashimoto, Y. and Daimon, S. and Iguchi, R. and Oikawa, Y. and Shen, K. and Sato, K. and Bossini, D. and Tabuchi, Y. and Satoh, T. and Hillebrands, B. and Bauer, G. E. W. and Johansen, T. H. and Kirilyuk, A. I. and Rasing, Th. and Saitoh, E.},
  title   = {{All-optical observation and reconstruction of spin wave dispersion}},
  journal = {Nature Communications},
  volume  = {8},
  pages   = {15859},
  year    = {2017},
  doi     = {10.1038/ncomms15859}
}

@article{Schlappa2024,
  author  = {Schlappa, J. and others},
  title   = {{The Heisenberg-RIXS instrument at the European XFEL}},
  journal = {Journal of Synchrotron Radiation},
  volume  = {32},
  pages   = {29--45},
  year    = {2025},
  doi     = {10.1107/S1600577524010890}
}

@article{Grimsditch1979,
  author  = {Grimsditch, M. and Malozemoff, A. and Brunsch, A.},
  title   = {{Standing Spin Waves Observed by Brillouin Scattering in Amorphous Metallic Fe$_{80}$B$_{20}$ Films}},
  journal = {Physical Review Letters},
  volume  = {43},
  pages   = {711},
  year    = {1979}
}

@article{Sandercock1979,
  author  = {Sandercock, J. R. and Wettling, W.},
  title   = {{Light scattering from surface and bulk thermal magnons in iron and nickel}},
  journal = {Journal of Applied Physics},
  volume  = {50},
  pages   = {7784--7789},
  year    = {1979}
}

@article{Kortright2013,
    title = {Resonant soft X-ray and extreme ultraviolet magnetic scattering in nanostructured magnetic materials: Fundamentals and directions},
    journal = {Journal of Electron Spectroscopy and Related Phenomena},
    volume = {189},
    pages = {178-186},
    year = {2013},
    issn = {0368-2048},
    doi = {https://doi.org/10.1016/j.elspec.2013.01.019},
    author = {Jeffrey B. Kortright}
}

@book{Eichler1986,
  author    = {Eichler, H. J. and Gunter, P. and Pohl, D. W.},
  title     = {{Laser-Induced Dynamic Gratings}},
  publisher = {Springer-Verlag},
  address   = {Berlin},
  year      = {1986}
}

@article{Bencivenga2019,
  author  = {Bencivenga, F. and Mincigrucci, R. and Capotondi, F. and Foglia, L. and Naumenko, D. and Maznev, A. A. and Pedersoli, E. and Simoncig, A. and Caporaletti, F. and Chiloyan, V. and Cucini, R. and Dallari, F. and Duncan, R. A. and Frazer, T. D. and Gaio, G. and Gessini, A. and Giannessi, L. and Huberman, S. and Kapteyn, H. and Knobloch, J. and Kurdi, G. and Mahne, N. and Manfredda, M. and Martinelli, A. and Murnane, M. and Principi, E. and Raimondi, L. and Spampinati, S. and Spezzani, C. and Trov{\`o}, M. and Zangrando, M. and Chen, G. and Monaco, G. and Nelson, K. A. and Masciovecchio, C.},
  title   = {{Nanoscale transient gratings excited and probed by extreme ultraviolet femtosecond pulses}},
  journal = {Science Advances},
  volume  = {5},
  pages   = {eaaw5805},
  year    = {2019}
}

@article{Rouxel2021,
  author  = {Rouxel, J. R. and Fainozzi, D. and Mankowsky, R. and R{\"o}sner, B. and Seniutinas, G. and Mincigrucci, R. and Catalini, S. and Foglia, L. and Cucini, R. and D{\"o}ring, F. and Kubec, A. and Koch, F. and Bencivenga, F. and Al Haddad, A. and Gessini, A. and Maznev, A. A. and Cirelli, C. and Gerber, S. and Pedrini, B. and Mancini, G. F. and Razzoli, E. and Burian, M. and Ueda, H. and Pamfilidis, G. and Ferrari, E. and Deng, Y. and Mozzanica, A. and Johnson, P. J. M. and Ozerov, D. and Izzo, M. G. and Bottari, C. and Arrell, C. and Divall, E. J. and Zerdane, S. and Sander, M. and Knopp, G. and Beaud, P. and Lemke, H. T. and Milne, C. J. and David, C. and Torre, R. and Chergui, M. and Nelson, K. A. and Masciovecchio, C. and Staub, U. and Patthey, L. and Svetina, C.},
  title   = {{Hard X-ray transient grating spectroscopy on bismuth germinate}},
  journal = {Nature Photonics},
  volume  = {15},
  pages   = {499--503},
  year    = {2021}
}

@article{Li2026,
  author  = {Li, H. and Wang, N. and Zhang, L. and Song, S. and Sun, Y. and Ng, M.-L. and Sato, T. and Hanlon, D. and Dahal, S. and Balcazar, M. D. and Esposito, V. and She, S. and Ornelas-Skarin, C. Caleb and Vila-Comamala, J. and David, C. and Berndt, N. and Miedaner, P. R. and Zhang, Z. and Ihme, M. and Trigo, M. and Nelson, K. A. and Hastings, J. B. and Maznev, A. A. and Foglia, L. and Teitelbaum, S. and Reis, D. A. and Zhu, D.},
  title   = {{Nanoscale Ultrafast Lattice Modulation with Hard X-ray Free Electron Laser}},
  journal = {Nature Physics},
  volume  = {22},
  pages   = {483},
  year    = {2026}
}

@article{Foglia2023,
  author  = {Foglia, L. and Mincigrucci, R. and Maznev, A. A. and Baldi, G. and Capotondi, F. and Caporaletti, F. and Comin, R. and De Angelis, D. and Duncan, R. A. and Fainozzi, D. and Kurdi, G. and Li, J. and Martinelli, A. and Masciovecchio, C. and Monaco, G. and Milloch, A. and Nelson, K. A. and Occhialini, C. A. and Pancaldi, M. and Pedersoli, E. and Pelli-Cresi, J. S. and Simoncig, A. and Travasso, F. and Wehinger, B. and Zanatta, M. and Bencivenga, F.},
  title   = {{Extreme ultraviolet transient gratings: A tool for nanoscale photoacoustics}},
  journal = {Photoacoustics},
  volume  = {29},
  pages   = {100453},
  year    = {2023}
}

@article{Ksenzov2021,
  author  = {Ksenzov, D. and Maznev, A. A. and Unikandanunni, V. and Bencivenga, F. and Capotondi, F. and Caretta, A. and Foglia, L. and Malvestuto, M. and Masciovecchio, C. and Mincigrucci, R. and Nelson, K. A. and Pancaldi, M. and Pedersoli, E. and Randolph, L. and Rahmann, H. and Urazhdin, S. and Bonetti, S. and Gutt, C.},
  title   = {{Nanoscale transient magnetization gratings created and probed by femtosecond extreme ultraviolet pulses}},
  journal = {Nano Letters},
  volume  = {21},
  pages   = {2905--2911},
  year    = {2021}
}

@article{Yao2022,
  author  = {Yao, K. and Steinbach, F. and Borchert, M. and Schick, D. and Engel, D. and Bencivenga, F. and Mincigrucci, R. and Foglia, L. and Pedersoli, E. and Angelis, D. D. and Pancaldi, M. and Wehinger, B. and Capotondi, F. and Masciovecchio, C. and Eisebitt, S. and Schmising, C. V. K.},
  title   = {{All-optical switching on the nanometer scale excited and probed with femtosecond extreme ultraviolet pulses}},
  journal = {Nano Letters},
  volume  = {22},
  pages   = {4452--4458},
  year    = {2022}
}

@article{vanKampen2002,
  author  = {van Kampen, M. and Jozsa, C. and Kohlhepp, J. T. and LeClair, P. and Lagae, L. and de Jonge, W. J. M. and Koopmans, B.},
  title   = {{All-optical probe of coherent spin waves}},
  journal = {Physical Review Letters},
  volume  = {88},
  pages   = {227201},
  year    = {2002}
}

@article{Cao2021,
  author  = {Cao, G. and Jiang, S. and Akerman, J. and Weissenrieder, J.},
  title   = {{Femtosecond laser driven precessing magnetic gratings}},
  journal = {Nanoscale},
  volume  = {13},
  pages   = {3746--3756},
  year    = {2021}
}

@article{Prieto2002,
  author  = {Prieto, J. E. and Heigl, F. and Krupin, O. and Kaindl, G. and Starke, K.},
  title   = {{Prediction of huge x-ray Faraday rotation at the Gd N$_{4,5}$ threshold}},
  journal = {Physical Review B},
  volume  = {66},
  pages   = {172408},
  year    = {2002}
}

@article{Man2017,
  author  = {Man, H. and Shi, Z. and Xu, G. and Xu, Y. and Chen, X. and Sullivan, S. and Zhou, J. and Xia, K. and Shi, J. and Dai, P.},
  title   = {{Direct observation of magnon-phonon coupling in yttrium iron garnet}},
  journal = {Physical Review B},
  volume  = {96},
  pages   = {100406},
  year    = {2017}
}

@article{Martinelli2022,
  author  = {Martinelli, L. and Betto, D. and Kummer, K. and Arpaia, R. and Braicovich, L. and Di Castro, D. and Brookes, N. B. and Sala, M. M. and Ghiringhelli, G.},
  title   = {{Fractional spin excitations in the infinite-layer cuprate CaCuO$_2$}},
  journal = {Physical Review X},
  volume  = {12},
  pages   = {021041},
  year    = {2022}
}

@article{Johnsen2023,
  author  = {Johnsen, P. C. and Ryan, S. A. and Gentry, C. and Grafov, A. and Kapteyn, H. and Murnane, M.},
  title   = {{A beamline for ultrafast extreme ultraviolet magneto-optical spectroscopy in reflection near the shotnoise limit}},
  journal = {Review of Scientific Instruments},
  volume  = {94},
  pages   = {033001},
  year    = {2023}
}

@article{Miedaner2026,
  author  = {Miedaner, P. R. and Maznev, A. A. and Biednov, M. and Deb, M. and Serrat, C. and Berndt, N. and Carrara, P. and Soncini, C. and Brioschi, M. and Ronchetti, D. and Benediktovitch, A. and Fainozzi, D. and Khatu, N. and Ferrari, E. and Vila-Comamala, J. and Zalden, P. and Gawelda, W. and Knoll, M. and Frankenberger, P. and Leroy, L. and Mamyrbayev, T. and Smolentsev, G. and Gerber, S. and Gessini, A. and Bencivenga, F. and Cucini, R. and Rossi, G. and Mincigrucci, R. and Paltanin, E. and Chergui, M. and Masciovecchio, C. and Bonetti, S. and Uemura, Y. and Huang, X. and Xu, H. and Alves Lima, F. and Ardana-Lamas, F. and Madsen, A. and Torre, R. and Ba{\~n}ares, L. and Szlachetko, J. and Blachucki, W. and Bargheer, M. and Feurer, T. and Engel, R. Y. and Beye, M. and David, C. and Staub, U. and Cannizzo, A. and Milne, C. and Nelson, K. A. and Svetina, C.},
  title   = {{Spin waves excited by hard x-ray transient gratings}},
  journal = {Physical Review Letters},
  volume  = {136},
  pages   = {176701},
  year    = {2026}
}

@article{Maznev2021,
  author  = {Maznev, A. A. and Mincigrucci, R. and Bencivenga, F. and Unikandanunni, V. and Capotondi, F. and Chen, G. and Ding, Z. and Duncan, R. A. and Foglia, L. and Izzo, M. G. and Masciovecchio, C. and Martinelli, A. and Monaco, G. and Pedersoli, E. and Bonetti, S. and Nelson, K. A.},
  title   = {{Generation and detection of 50 GHz surface acoustic waves by extreme ultraviolet pulses}},
  journal = {Applied Physics Letters},
  volume  = {119},
  pages   = {044102},
  year    = {2021}
}

@article{Weigand2022,
  author  = {Weigand, M. and Wintz, S. and Gr{\"a}fe, J. and Noske, M. and Stoll, H. and Van Waeyenberge, B. and Sch{\"u}tz, G.},
  title   = {{TimeMaxyne: A Shot-Noise Limited, Time-Resolved Pump-and-Probe Acquisition System Capable of 50 GHz Frequencies for Synchrotron-Based X-ray Microscopy}},
  journal = {Crystals},
  volume  = {12},
  pages   = {1029},
  year    = {2022},
  doi     = {10.3390/cryst12081029}
}

@article{Finizio2026,
  author  = {Finizio, S. and Butcher, T. A. and Wintz, S. and Weigand, M. and Raabe, J.},
  title   = {{Scanning through space and time: past, present, and future of time-resolved scanning transmission soft X-ray microscopy}},
  journal = {Journal of Physics D: Applied Physics},
  year    = {2026},
  doi     = {10.1088/1361-6463/aea628},
  volume = {59},
  number = {38},
}

@article{Schutz1987,
  author  = {Sch{\"u}tz, G. and Wagner, W. and Wilhelm, W. and Kienle, P. and Zeller, R. and Frahm, R. and Materlik, G.},
  title   = {{Absorption of circularly polarized x rays in iron}},
  journal = {Physical Review Letters},
  volume  = {58},
  pages   = {737--740},
  year    = {1987},
  doi     = {10.1103/PhysRevLett.58.737}
}

@article{Yamamoto2025,
  author  = {Yamamoto, R. and Turnbull, L. A. and Schmidt, M. and Corsaletti Filho, J. C. and Binger, H. J. and Di Pietro Mart{\'\i}nez, M. and Weigand, M. and Finizio, S. and Prots, Y. and Ferguson, G. M. and Vool, U. and Wintz, S. and Donnelly, C.},
  title   = {{Altermagnetic nanotextures revealed in bulk MnTe}},
  journal = {Physical Review Applied},
  volume  = {24},
  pages   = {034037},
  year    = {2025},
  doi     = {10.1103/dp7v-qszq}
}

@misc{Yamamoto2026,
  author  = {Yamamoto, R. and Mayr, S. and Hariki, A. and Finizio, S. and Sakurai, K. and Weschke, E. and Litzius, K. and Birch, M. T. and Turnbull, L. A. and Zhakina, E. and Di Pietro Mart{\'\i}nez, M. and Reuteler, J. and Schulz, F. and Weigand, M. and Raabe, J. and Sch{\"u}tz, G. and Arekapudi, S. S. P. K. and Hellwig, O. and Campos, W. H. and {\v{S}}mejkal, L. and Kune{\v{s}}, J. and Donnelly, C. and Wintz, S.},
  title   = {{Nanoscale imaging of spin textures with locally varying altermagnetic response in $\alpha$-Fe$_2$O$_3$}},
  howpublished = {arXiv:2603.09934},
  year    = {2026},
  doi     = {10.48550/arXiv.2603.09934}
}

@article{Birch2026,
  author  = {Birch, M. T. and Wintz, S. and Sun, Y. and Kikkawa, A. and Weigand, M. and Arima, T. and Tokura, Y.},
  title   = {{X-ray imaging of antiferromagnetic cluster octupole domains in Mn$_3$Sn}},
  journal = {Physical Review B},
  volume  = {114},
  pages   = {094409},
  year    = {2026},
  doi     = {10.1103/fq34-r263}
}

@article{vanderLaan1986,
  author  = {van der Laan, G. and Thole, B. T. and Sawatzky, G. A. and Goedkoop, J. B. and Fuggle, J. C. and Esteva, J.-M. and Karnatak, R. and Remeika, J. P. and Dabkowska, H. A.},
  title   = {{Experimental proof of magnetic x-ray dichroism}},
  journal = {Physical Review B},
  volume  = {34},
  pages   = {6529--6531},
  year    = {1986},
  doi     = {10.1103/PhysRevB.34.6529}
}

@article{Koraltan2025,
  author  = {Koraltan, S. and Porrati, F. and Kraft, R. and Barth, S. and Weigand, M. and Abert, C. and Suess, D. and Huth, M. and Wintz, S.},
  title   = {{Reconfigurable 3D Magnetic Nanoarchitectures}},
  journal = {Advanced Functional Materials},
  volume  = {35},
  pages   = {e15338},
  year    = {2025},
  doi     = {10.1002/adfm.202515338}
}

@article{Xu2026,
  author  = {Xu, M. and Deenen, A. J. M. and Guo, H. and Morales-Fern{\'a}ndez, P. and Wintz, S. and Zhakina, E. and Weigand, M. and Donnelly, C. and Grundler, D.},
  title   = {{Geometry-induced spin chirality in a non-chiral ferromagnet at zero field}},
  journal = {Nature Nanotechnology},
  volume  = {21},
  pages   = {58--64},
  year    = {2026},
  doi     = {10.1038/s41565-025-02055-3}
}

@article{VanWaeyenberge2006,
  author  = {Van Waeyenberge, B. and Puzic, A. and Stoll, H. and Chou, K. W. and Tyliszczak, T. and Hertel, R. and F{\"a}hnle, M. and Br{\"u}ckl, H. and Rott, K. and Reiss, G. and Neudecker, I. and Weiss, D. and Back, C. H. and Sch{\"u}tz, G.},
  title   = {{Magnetic vortex core reversal by excitation with short bursts of an alternating field}},
  journal = {Nature},
  volume  = {444},
  pages   = {461--464},
  year    = {2006},
  doi     = {10.1038/nature05240}
}

@article{Acremann2006,
  author  = {Acremann, Y. and Strachan, J. P. and Chembrolu, V. and Andrews, S. D. and Tyliszczak, T. and Katine, J. A. and Carey, M. J. and Clemens, B. M. and Siegmann, H. C. and St{\"o}hr, J.},
  title   = {{Time-Resolved Imaging of Spin Transfer Switching: Beyond the Macrospin Concept}},
  journal = {Physical Review Letters},
  volume  = {96},
  pages   = {217202},
  year    = {2006},
  doi     = {10.1103/PhysRevLett.96.217202}
}

@article{Kammerer2011,
  author  = {Kammerer, M. and Weigand, M. and Curcic, M. and Noske, M. and Sproll, M. and Vansteenkiste, A. and Van Waeyenberge, B. and Stoll, H. and Woltersdorf, G. and Back, C. H. and Schuetz, G.},
  title   = {{Magnetic vortex core reversal by excitation of spin waves}},
  journal = {Nature Communications},
  volume  = {2},
  pages   = {279},
  year    = {2011},
  doi     = {10.1038/ncomms1277}
}

@article{Cheng2012,
  author  = {Cheng, C. and Bailey, W. E.},
  title   = {{Sub-micron mapping of GHz magnetic susceptibility using scanning transmission x-ray microscopy}},
  journal = {Applied Physics Letters},
  volume  = {101},
  pages   = {182407},
  year    = {2012},
  doi     = {10.1063/1.4765663}
}

@article{Bonetti2015,
  author  = {Bonetti, S. and Kukreja, R. and Chen, Z. and Maci{\`a}, F. and Hern{\'a}ndez, J. M. and Eklund, A. and Backes, D. and Frisch, J. and Katine, J. and Malm, G. and Urazhdin, S. and Kent, A. D. and St{\"o}hr, J. and Ohldag, H. and D{\"u}rr, H. A.},
  title   = {{Direct observation and imaging of a spin-wave soliton with p-like symmetry}},
  journal = {Nature Communications},
  volume  = {6},
  pages   = {8889},
  year    = {2015},
  doi     = {10.1038/ncomms9889}
}

@article{Stoll2004,
  author  = {Stoll, H. and Puzic, A. and van Waeyenberge, B. and Fischer, P. and Raabe, J. and Buess, M. and Haug, T. and H{\"o}llinger, R. and Back, C. and Weiss, D. and Denbeaux, G.},
  title   = {{High-resolution imaging of fast magnetization dynamics in magnetic nanostructures}},
  journal = {Applied Physics Letters},
  volume  = {84},
  pages   = {3328--3330},
  year    = {2004},
  doi     = {10.1063/1.1723698}
}

@article{Choe2004,
  author  = {Choe, S.-B. and Acremann, Y. and Scholl, A. and Bauer, A. and Doran, A. and St{\"o}hr, J. and Padmore, H. A.},
  title   = {{Vortex Core-Driven Magnetization Dynamics}},
  journal = {Science},
  volume  = {304},
  pages   = {420--422},
  year    = {2004},
  doi     = {10.1126/science.1095068}
}

@article{Buttner2015,
  author  = {B{\"u}ttner, F. and Moutafis, C. and Schneider, M. and Kr{\"u}ger, B. and G{\"u}nther, C. M. and Geilhufe, J. and von Korff Schmising, C. and Mohanty, J. and Pfau, B. and Schaffert, S. and Bisig, A. and Foerster, M. and Schulz, T. and Vaz, C. A. F. and Franken, J. H. and Swagten, H. J. M. and Kl{\"a}ui, M. and Eisebitt, S.},
  title   = {{Dynamics and inertia of skyrmionic spin structures}},
  journal = {Nature Physics},
  volume  = {11},
  pages   = {225--228},
  year    = {2015},
  doi     = {10.1038/nphys3234}
}

@article{Donnelly2020,
  author  = {Donnelly, C. and Finizio, S. and Gliga, S. and Holler, M. and Hrabec, A. and Odstr{\v{c}}il, M. and Mayr, S. and Scagnoli, V. and Heyderman, L. J. and Guizar-Sicairos, M. and Raabe, J.},
  title   = {{Time-resolved imaging of three-dimensional nanoscale magnetization dynamics}},
  journal = {Nature Nanotechnology},
  volume  = {15},
  pages   = {356--360},
  year    = {2020},
  doi     = {10.1038/s41565-020-0649-x}
}

@article{Butcher2025,
  author  = {Butcher, T. A. and Phillips, N. W. and Levitan, A. L. and Weigand, M. and Wintz, S. and Raabe, J. and Finizio, S.},
  title   = {{Nanoscale domain-wall dynamics in micromagnetic structures with weak perpendicular anisotropy}},
  journal = {Physical Review B},
  volume  = {111},
  pages   = {L220409},
  year    = {2025},
  doi     = {10.1103/2y4r-my27}
}

@article{Wintz2016,
  author  = {Wintz, S. and Tiberkevich, V. and Weigand, M. and Raabe, J. and Lindner, J. and Erbe, A. and Slavin, A. and Fassbender, J.},
  title   = {{Magnetic vortex cores as tunable spin-wave emitters}},
  journal = {Nature Nanotechnology},
  volume  = {11},
  pages   = {948--953},
  year    = {2016},
  doi     = {10.1038/nnano.2016.117}
}

@article{Dieterle2019,
  author  = {Dieterle, G. and F{\"o}rster, J. and Stoll, H. and Semisalova, A. S. and Finizio, S. and Gangwar, A. and Weigand, M. and Noske, M. and F{\"a}hnle, M. and Bykova, I. and Gr{\"a}fe, J. and Bozhko, D. A. and Musiienko-Shmarova, H. Yu. and Tiberkevich, V. and Slavin, A. N. and Back, C. H. and Raabe, J. and Sch{\"u}tz, G. and Wintz, S.},
  title   = {{Coherent Excitation of Heterosymmetric Spin Waves with Ultrashort Wavelengths}},
  journal = {Physical Review Letters},
  volume  = {122},
  pages   = {117202},
  year    = {2019},
  doi     = {10.1103/PhysRevLett.122.117202}
}

@article{Forster2019prb,
  author  = {F{\"o}rster, J. and Gr{\"a}fe, J. and Bailey, J. and Finizio, S. and Tr{\"a}ger, N. and Gro{\ss}, F. and Mayr, S. and Stoll, H. and Dubs, C. and Surzhenko, A. B. and Liebing, N. and Woltersdorf, G. and Raabe, J. and Weigand, M. and Sch{\"u}tz, G. and Wintz, S.},
  title   = {{Direct observation of coherent magnons with suboptical wavelengths in a single-crystalline ferrimagnetic insulator}},
  journal = {Physical Review B},
  volume  = {100},
  pages   = {214416},
  year    = {2019},
  doi     = {10.1103/PhysRevB.100.214416}
}

@article{Baumgaertl2020,
  author  = {Baumgaertl, K. and Gr{\"a}fe, J. and Che, P. and Mucchietto, A. and F{\"o}rster, J. and Tr{\"a}ger, N. and Bechtel, M. and Weigand, M. and Sch{\"u}tz, G. and Grundler, D.},
  title   = {{Nanoimaging of Ultrashort Magnon Emission by Ferromagnetic Grating Couplers at GHz Frequencies}},
  journal = {Nano Letters},
  volume  = {20},
  pages   = {7281--7286},
  year    = {2020},
  doi     = {10.1021/acs.nanolett.0c02645}
}

@article{Mayr2021,
  author  = {Mayr, S. and Flaj{\v{s}}man, L. and Finizio, S. and Hrabec, A. and Weigand, M. and F{\"o}rster, J. and Stoll, H. and Heyderman, L. J. and Urb{\'a}nek, M. and Wintz, S. and Raabe, J.},
  title   = {{Spin-Wave Emission from Vortex Cores under Static Magnetic Bias Fields}},
  journal = {Nano Letters},
  volume  = {21},
  pages   = {1584--1590},
  year    = {2021},
  doi     = {10.1021/acs.nanolett.0c03740}
}

@article{Gallardo2024,
  author  = {Gallardo, R. and Weigand, M. and Schultheiss, K. and Kakay, A. and Mattheis, R. and Raabe, J. and Sch{\"u}tz, G. and Deac, A. and Lindner, J. and Wintz, S.},
  title   = {{Coherent Magnons with Giant Nonreciprocity at Nanoscale Wavelengths}},
  journal = {ACS Nano},
  volume  = {18},
  pages   = {5249},
  year    = {2024},
  doi     = {10.1021/acsnano.3c08390}
}

@article{Koraltan2024,
  author  = {Koraltan, S. and Schultheiss, K. and Bruckner, F. and Weigand, M. and Abert, C. and Suess, D. and Wintz, S.},
  title   = {{Steerable current-driven emission of spin waves in magnetic vortex pairs}},
  journal = {Science Advances},
  volume  = {10},
  pages   = {eado8635},
  year    = {2024},
  doi     = {10.1126/sciadv.ado8635}
}

@article{Behncke2018,
  author  = {Behncke, C. and Adolff, C. F. and Lenzing, N. and H{\"a}nze, M. and Schulte, B. and Weigand, M. and Sch{\"u}tz, G. and Meier, G.},
  title   = {{Spin-wave interference in magnetic vortex stacks}},
  journal = {Communications Physics},
  volume  = {1},
  pages   = {50},
  year    = {2018},
  doi     = {10.1038/s42005-018-0052-1}
}

@article{Albisetti2018,
  author  = {Albisetti, E. and Petti, D. and Sala, G. and Silvani, R. and Tacchi, S. and Finizio, S. and Wintz, S. and Cal{\`o}, A. and Zheng, X. and Raabe, J. and Riedo, E. and Bertacco, R.},
  title   = {{Nanoscale spin-wave circuits based on engineered reconfigurable spin-textures}},
  journal = {Communications Physics},
  volume  = {1},
  pages   = {56},
  year    = {2018},
  doi     = {10.1038/s42005-018-0056-x}
}

@article{Che2025,
  author  = {Che, P. and Deenen, A. J. M. and Mucchietto, A. and Gr{\"a}fe, J. and Heigl, M. and Baumgaertl, K. and Weigand, M. and Bechtel, M. and Koraltan, S. and Sch{\"u}tz, G. and Suess, D. and Albrecht, M. and Grundler, D.},
  title   = {{Nonreciprocal Spin Waves in Nanoscale Hybrid N{\'e}el--Bloch--N{\'e}el Domain Walls Detected by Scanning X-Ray Microscopy in Perpendicular Magnetic Anisotropic Fe/Gd Multilayers}},
  journal = {Advanced Materials},
  volume  = {37},
  pages   = {e08181},
  year    = {2025},
  doi     = {10.1002/adma.202508181}
}

@article{Albisetti2020,
  author  = {Albisetti, E. and Tacchi, S. and Silvani, R. and Scaramuzzi, G. and Finizio, S. and Wintz, S. and Rinaldi, C. and Cantoni, M. and Raabe, J. and Carlotti, G. and Bertacco, R. and Riedo, E. and Petti, D.},
  title   = {{Optically Inspired Nanomagnonics with Nonreciprocal Spin Waves in Synthetic Antiferromagnets}},
  journal = {Advanced Materials},
  volume  = {32},
  pages   = {1906439},
  year    = {2020},
  doi     = {10.1002/adma.201906439}
}

@article{Talapatra2023,
  author  = {Talapatra, A. and Qin, H. and Schulz, F. and Yao, L. and Flaj{\v{s}}man, L. and Weigand, M. and Wintz, S. and van Dijken, S.},
  title   = {{Imaging of short-wavelength spin waves in a nanometer-thick YIG/Co bilayer}},
  journal = {Applied Physics Letters},
  volume  = {122},
  pages   = {202404},
  year    = {2023},
  doi     = {10.1063/5.0149583}
}

@article{Gross2019,
  author  = {Gro{\ss}, F. and Tr{\"a}ger, N. and F{\"o}rster, J. and Weigand, M. and Sch{\"u}tz, G. and Gr{\"a}fe, J.},
  title   = {{Nanoscale detection of spin wave deflection angles in permalloy}},
  journal = {Applied Physics Letters},
  volume  = {114},
  pages   = {012406},
  year    = {2019},
  doi     = {10.1063/1.5074169}
}

@article{Forster2019jap,
  author  = {F{\"o}rster, J. and Wintz, S. and Bailey, J. and Finizio, S. and Josten, E. and Dubs, C. and Bozhko, D. A. and Stoll, H. and Dieterle, G. and Tr{\"a}ger, N. and Raabe, J. and Slavin, A. N. and Weigand, M. and Gr{\"a}fe, J. and Sch{\"u}tz, G.},
  title   = {{Nanoscale X-ray imaging of spin dynamics in yttrium iron garnet}},
  journal = {Journal of Applied Physics},
  volume  = {126},
  pages   = {173909},
  year    = {2019},
  doi     = {10.1063/1.5121013}
}

@article{Schulz2023aip,
  author  = {Schulz, F. and Gro{\ss}, F. and F{\"o}rster, J. and Mayr, S. and Weigand, M. and Goering, E. and Gr{\"a}fe, J. and Sch{\"u}tz, G. and Wintz, S.},
  title   = {{Realization of a magnonic analog adder with frequency-division multiplexing}},
  journal = {AIP Advances},
  volume  = {13},
  pages   = {015115},
  year    = {2023},
  doi     = {10.1063/5.0120826}
}

@article{Greil2025,
  author  = {Greil, J. and Kiechle, M. and Papp, A. and Neumann, P. and Kov{\'a}cs, Z. and Volk, J. and Schulz, F. and Wintz, S. and Weigand, M. and Csaba, G. and Becherer, M.},
  title   = {{The effect of Ga-ion irradiation on sub-micron-wavelength spin waves in yttrium-iron-garnet films}},
  journal = {Nanotechnology},
  volume  = {36},
  pages   = {135301},
  year    = {2025},
  doi     = {10.1088/1361-6528/adad7d}
}

@misc{Wojewoda2026,
  author  = {Wojewoda, O. and Mayr, S. and Gross, M. J. and Kl{\'\i}ma, J. and Panda, J. and Kr{\v{c}}ma, J. and Holobr{\'a}dek, J. and Dav{\'\i}dkov{\'a}, K. and Chumak, A. V. and Gerasimchuk, I. and Verba, R. and Pirro, P. and Weigand, M. and Finizio, S. and Lindner, M. and Dubs, C. and Wang, Q. and Wintz, S. and Ross, C. A. and Urb{\'a}nek, M.},
  title   = {{Ultrafast formation of a large dynamic magnetic soliton}},
  howpublished = {arXiv:2606.00849},
  year    = {2026},
  doi     = {10.48550/arXiv.2606.00849}
}

@article{Schulz2023nl,
  author  = {Schulz, F. and Litzius, K. and Powalla, L. and Birch, M. T. and Gallardo, R. A. and Satheesh, S. and Weigand, M. and Scholz, T. and Lotsch, B. V. and Sch{\"u}tz, G. and Burghard, M. and Wintz, S.},
  title   = {{Direct Observation of Propagating Spin Waves in the 2D van der Waals Ferromagnet Fe$_5$GeTe$_2$}},
  journal = {Nano Letters},
  volume  = {23},
  pages   = {10126--10131},
  year    = {2023},
  doi     = {10.1021/acs.nanolett.3c02212}
}

@article{Gonzalez2026,
  author  = {Gonz{\'a}lez, V. H. and Schulz, F. and Behera, N. and Ahlberg, M. and Kumar, A. and Frisk, A. and Gro{\ss}, F. and Ilse, S. E. and Wittrock, S. and Weigand, M. and Sch{\"u}tz, G. and {\AA}kerman, J. and Wintz, S.},
  title   = {{Direct Observation of Propagating Spin Waves in a Spin Hall Nano-Oscillator}},
  journal = {Advanced Materials},
  volume  = {38},
  pages   = {e74547},
  year    = {2026},
  doi     = {10.1002/adma.74547}
}

@article{Korber2021,
  author  = {K{\"o}rber, L. and Zimmermann, M. and Wintz, S. and Finizio, S. and Kronseder, M. and Bougeard, D. and Dirnberger, F. and Weigand, M. and Raabe, J. and Ot{\'a}lora, J. A. and Schultheiss, H. and Josten, E. and Lindner, J. and K{\'e}zsm{\'a}rki, I. and Back, C. H. and K{\'a}kay, A.},
  title   = {{Symmetry and curvature effects on spin waves in vortex-state hexagonal nanotubes}},
  journal = {Physical Review B},
  volume  = {104},
  pages   = {184429},
  year    = {2021},
  doi     = {10.1103/PhysRevB.104.184429}
}

@article{MoralesFernandez2026,
  author  = {Morales-Fern{\'a}ndez, P. and Konstantinos-Douveas, I. and Abert, C. and Zhakina, E. and Ruiz-G{\'o}mez, S. and Fern{\'a}ndez-Gonz{\'a}lez, C. and Turnbull, L. A. and Wintz, S. and K{\"o}nig, M. and Finizio, S. and Weigand, M. and Leo, N. and Suess, D. and Hierro-Rodr{\'\i}guez, A. and Fern{\'a}ndez-Pacheco, A. and Donnelly, C.},
  title   = {{Resonant Domain Wall Dynamics in a Three-Dimensional Magnetic Nano Double Helix}},
  journal = {Advanced Materials},
  pages   = {e00004},
  year    = {2026},
  doi     = {10.1002/adma.202600004}
}

@article{Trager2020pssr,
  author  = {Tr{\"a}ger, N. and Gruszecki, P. and Lisiecki, F. and F{\"o}rster, J. and Weigand, M. and Wintz, S. and Stoll, H. and G{\l}owi{\'n}ski, H. and Ku{\'s}wik, P. and Krawczyk, M. and Gr{\"a}fe, J.},
  title   = {{Direct Imaging of High-Frequency Multimode Spin Wave Propagation in Cobalt-Iron Waveguides Using X-Ray Microscopy beyond 10 GHz}},
  journal = {physica status solidi (RRL) -- Rapid Research Letters},
  volume  = {14},
  pages   = {2000373},
  year    = {2020},
  doi     = {10.1002/pssr.202000373}
}

@article{Trager2021,
  author  = {Tr{\"a}ger, N. and Gruszecki, P. and Lisiecki, F. and Gross, F. and F{\"o}rster, J. and Weigand, M. and G{\l}owi{\'n}ski, H. and Ku{\'s}wik, P. and Dubowik, J. and Sch{\"u}tz, G. and Krawczyk, M. and Gr{\"a}fe, J.},
  title   = {{Real-Space Observation of Magnon Interaction with Driven Space-Time Crystals}},
  journal = {Physical Review Letters},
  volume  = {126},
  pages   = {057201},
  year    = {2021},
  doi     = {10.1103/PhysRevLett.126.057201}
}

@article{Pile2022,
  author  = {Pile, S. and Stienen, S. and Lenz, K. and Narkowicz, R. and Wintz, S. and F{\"o}rster, J. and Mayr, S. and Buchner, M. and Weigand, M. and Ney, V. and Lindner, J. and Ney, A.},
  title   = {{Nonstationary spin waves in a single rectangular permalloy microstrip under uniform magnetic excitation}},
  journal = {Physical Review B},
  volume  = {105},
  pages   = {094415},
  year    = {2022},
  doi     = {10.1103/PhysRevB.105.094415}
}

@article{Feggeler2023,
  author  = {Feggeler, T. and Lill, J. and G{\"u}nzing, D. and Meckenstock, R. and Spoddig, D. and Efremova, M. V. and Wintz, S. and Weigand, M. and Zingsem, B. W. and Farle, M. and Wende, H. and Ollefs, K. J. and Ohldag, H.},
  title   = {{Spatially-resolved dynamic sampling of different phasic magnetic resonances of nanoparticle ensembles in a magnetotactic bacterium Magnetospirillum magnetotacticum}},
  journal = {New Journal of Physics},
  volume  = {25},
  pages   = {043010},
  year    = {2023},
  doi     = {10.1088/1367-2630/acc81f}
}

@article{Talmelli2020,
  author  = {Talmelli, G. and Devolder, T. and Tr{\"a}ger, N. and F{\"o}rster, J. and Wintz, S. and Weigand, M. and Stoll, H. and Heyns, M. and Sch{\"u}tz, G. and Radu, I. P. and Gr{\"a}fe, J. and Ciubotaru, F. and Adelmann, C.},
  title   = {{Reconfigurable submicrometer spin-wave majority gate with electrical transducers}},
  journal = {Science Advances},
  volume  = {6},
  pages   = {eabb4042},
  year    = {2020},
  doi     = {10.1126/sciadv.abb4042}
}

@article{Lisiecki2019,
  author  = {Lisiecki, F. and Rych{\l}y, J. and Ku{\'s}wik, P. and G{\l}owi{\'n}ski, H. and K{\l}os, J. W. and Gro{\ss}, F. and Tr{\"a}ger, N. and Bykova, I. and Weigand, M. and Zelent, M. and Goering, E. J. and Sch{\"u}tz, G. and Krawczyk, M. and Stobiecki, F. and Dubowik, J. and Gr{\"a}fe, J.},
  title   = {{Magnons in a Quasicrystal: Propagation, Extinction, and Localization of Spin Waves in Fibonacci Structures}},
  journal = {Physical Review Applied},
  volume  = {11},
  pages   = {054061},
  year    = {2019},
  doi     = {10.1103/PhysRevApplied.11.054061}
}

@article{Grafe2020,
  author  = {Gr{\"a}fe, J. and Gruszecki, P. and Zelent, M. and Decker, M. and Keskinbora, K. and Noske, M. and Gawronski, P. and Stoll, H. and Weigand, M. and Krawczyk, M. and Back, C. H. and Goering, E. J. and Sch{\"u}tz, G.},
  title   = {{Direct observation of spin-wave focusing by a Fresnel lens}},
  journal = {Physical Review B},
  volume  = {102},
  pages   = {024420},
  year    = {2020},
  doi     = {10.1103/PhysRevB.102.024420}
}

@article{Blank2023,
  title = {Empowering Control of Antiferromagnets by THz-Induced Spin Coherence},
  author = {Blank, T. G. H. and Grishunin, K. A. and Ivanov, B. A. and Mashkovich, E. A. and Afanasiev, D. and Kimel, A. V.},
  journal = {Phys. Rev. Lett.},
  volume = {131},
  issue = {9},
  pages = {096701},
  numpages = {6},
  year = {2023},
  month = {Aug},
  publisher = {American Physical Society},
  doi = {10.1103/PhysRevLett.131.096701},
}

@article{Bittencourt2026,
  title = {Quantum Impurity Sensing of Altermagnetic Order},
  author = {Bittencourt, V. A. S. V. and Hosseinabadi, Hossein and Sinova, Jairo and {\ifmmode \check{S}\else \v{S}\fi{}}mejkal, Libor and Marino, Jamir},
  journal = {Phys. Rev. Lett.},
  volume = {136},
  issue = {14},
  pages = {146701},
  numpages = {8},
  year = {2026},
  month = {Apr},
  publisher = {American Physical Society},
  doi = {10.1103/2ppn-kvjv}
}

\end{document}